\documentclass{article}

\usepackage[T1]{fontenc}
\usepackage{lmodern}
\usepackage{optidef} % for linear programming

\usepackage{bookmark}

\usepackage{microtype}
\usepackage{graphicx}
\usepackage{subcaption}

\usepackage{booktabs} % for professional tables
\usepackage{multirow}
\usepackage{tabularx}
\usepackage{makecell}

\usepackage{hyperref}

\usepackage{xcolor}
\usepackage{listings}

\definecolor{commentgreen}{rgb}{0, 0.5, 0}
\usepackage[accepted]{icml2026}

\usepackage{amsmath}
\usepackage{amssymb}
\usepackage{mathtools}
\usepackage{amsthm}

\usepackage{xspace}

\usepackage{tikz}

\usepackage[inline]{enumitem}
\newenvironment{denseitemize}{
    \begin{itemize}[topsep=2pt, partopsep=0pt, leftmargin=1.5em]
        \setlength{\itemsep}{2pt}
        \setlength{\parskip}{0pt}
        \setlength{\parsep}{0pt}
    }{\end{itemize}}

\newenvironment{denseenum}{
    \begin{enumerate}[topsep=2pt, partopsep=0pt, leftmargin=1.5em]
        \setlength{\itemsep}{2pt}
        \setlength{\parskip}{0pt}
        \setlength{\parsep}{0pt}
    }{\end{enumerate}}

\def\ie{{i.e.\xspace}}
\def\eg{{e.g.\xspace}}

\def\etc{etc.\xspace}

\theoremstyle{plain}

\theoremstyle{definition}

\theoremstyle{remark}

\usepackage[disable,textsize=tiny]{todonotes}

\newcommand{\name}{Cornstarch\xspace}
\newcommand{\parabf}[1]{\noindent\textbf{#1}\xspace}

\icmltitlerunning{Efficient Distributed MLLM Training with \name}

\begin{document}

\twocolumn[
  \icmltitle{Efficient Distributed MLLM Training with \name}



  \begin{icmlauthorlist}
    \icmlauthor{Insu Jang}{sch}
    \icmlauthor{Runyu Lu}{sch}
    \icmlauthor{Nikhil Bansal}{sch}
    \icmlauthor{Ang Chen}{sch}
    \icmlauthor{Mosharaf Chowdhury}{sch}
  \end{icmlauthorlist}

  \icmlaffiliation{sch}{University of Michigan}
  
  \icmlcorrespondingauthor{Insu Jang}{insujang@umich.edu}

  \icmlkeywords{Machine Learning, ICML, ML training system, multimodal LLM, large multimodal model training, distributed training, pipeline parallelism, context parallelism}

  \vskip 0.3in
]



\printAffiliationsAndNotice{}  % no special notice (required even if empty)

\begin{abstract} 
Multimodal large language models (MLLMs) extend the capabilities of large language models (LLMs) by combining heterogeneous model architectures to handle diverse modalities like images and audio.
However, this inherent heterogeneity in MLLM model structure and data types makes makeshift extensions to existing LLM training frameworks unsuitable for efficient MLLM training, especially in distributed training.
% While there are a few works that have attempted to address the heterogeneity in MLLM training, their approaches are limited to only superficially considering the characteristics of MLLMs.

In this paper, we present \name, an efficient distributed MLLM training framework that contemplates MLLM's unique characteristics in both model and data parallelization.
\name introduces frozen-aware pipeline parallelism and workload-balanced context parallelism to improve MLLM training throughput.
Our extensive evaluation shows that \name outperforms state-of-the-art solutions by $2.26\times$ on average in terms of MLLM training throughput.

\name is an open-source project available on Github~\footnote{https://github.com/cornstarch-org/Cornstarch}.
\end{abstract}

\section{Introduction}
\label{sec:introduction}

% MLLMs are important
Multimodal large language models (MLLMs) aim to extend LLMs' reasoning capabilities to perform complex tasks across various data modalities, such as images and audio~\cite{qwen2vl-arxiv24,videollava-arxiv24,vila-cvpr24,internvl-cvpr24,vlm-acl21,whisper-icml23,qwen2audio-arxiv24,llama3-arxiv24,ll3da-cvpr24,languagebind-iclr24,mm-llms-acl24,cambrian-corr24,llava-neurips23,llavanext-cvpr24,phi4-arxiv24}.
% For instance, MLLMs are used in healthcare to analyze medical images and patient records, aiding in accurate diagnoses~\cite{mllm-healthcare-jmlr23,mllm-healthcare-jmlr24,mllm-healthcare-npj23}. 
% In robotics, they process visual and auditory inputs, enabling robots to interact with their environment~\cite{mllm-robotics-crl23,mllm-robotics-icra23}.
% As the volume of multimodal data continues to grow, the importance of MLLMs will only increase.
% MLLMs are different. Highlight the unique characteristics
% The way of training MLLMs differs from LLM training.
While MLLMs can be trained from scratch like traditional LLMs, they are more commonly constructed by integrating \textit{pretrained} modality-specific encoders with language models~\cite{llavanext-cvpr24,deepfusion-survey24}.
Each modality's input is first processed by its corresponding encoder, then projected into a shared text embedding space through learnable projection layers, and finally processed by the LLM with text tokens.
% Unlike traditional LLM training, where all parameters are typically updated, MLLM training exhibits greater variability: aside from the learnable projection layers, the modality encoders and the LLM may be frozen or fine-tuned depending on the specific training strategy.
% This strategy reduces computational overhead while leveraging the representational capacity of the pretrained components. 

% \insu{Make GraphPipe SOTA existing work. Set two baselines: LLM 4D and GraphPipe (or just GraphPipe + 4D). Introduce baseline in Section 3 so that we add ours on top of them. Thus modality parallelism can still be introduced.}

The larger size of MLLMs and the need for more data processing power make distributed MLLM training essential.
However, heterogeneity in model and data makes balanced MLLM workload distribution more challenging and tackled by recent works~\cite{distmm-nsdi24,disttrain-sigcomm25,optimus-atc25,graphpipe-asplos25}.
% However, heterogeneity in model and data, and differences between MLLM and LLM training processes make balanced MLLM workload distribution more challenging.
% Indeed, several recent works have attempted to address various aspects of the model and data heterogeneity problem~\cite{distmm-nsdi24,disttrain-sigcomm25,optimus-atc25,graphpipe-asplos25}.
%  and distribute non-sequential execution flow of parallel modality encoders~\cite{graphpipe-asplos25}.
We observe that beyond the first-order disparities in model and data heterogeneity, there are two additional MLLM-specific distributed training challenges that have significant performance implications.
First, MLLM training with \emph{frozen} versus \emph{trainable} models results in different computational costs across modules.
Model partitioning strategies that do not account for the frozen status of components can lead to suboptimal performance.
Second, cross-modality interactions introduce non-causal attention patterns to enable more precise computation of their relationships~\cite{flexattention-mlsys25,flashmask-iclr25}.
While distributing causal attention patterns in LLMs has been extensively studied~\cite{contextparallel-mlsys25,wlbllm-osdi25,ringattention-iclr24}, \emph{efficient distribution of non-causal attention patterns} remains an open challenge.

% Cornstarch
In this paper, we introduce \name, an efficient distributed MLLM training framework.
\name transcends the first-order model and data heterogeneity-aware parallelization and uncovers latent higher-order heterogeneities in MLLMs that have not been considered in previous works~\cite{distmm-nsdi24,disttrain-sigcomm25,optimus-atc25,graphpipe-asplos25}.
% We identify that elusive MLLM characteristics have not been considered in previous works~\cite{distmm-nsdi24,disttrain-sigcomm25,optimus-atc25,graphpipe-asplos25} and incorporates them into the parallelization process.

\parabf{Frozen status-aware pipeline parallelism (\S\ref{sec:pipeline_parallelism}).}
\name introduces a way to consider the frozen status of MLLM components in model partitioning.
We observe that the frozen status of MLLM components can significantly affect the pipeline stage balancing.
Existing MLLM approaches do not consider the frozen status in model partitioning~\cite{distmm-nsdi24,disttrain-sigcomm25,optimus-atc25}.
Even with profiler-based automated approaches that actually measure the backward pass time~\cite{pipedream-icml21,galvatron-vldb22}, they cannot account for different computational costs of modules due to the frozen status.
We precisely compute the backward pass time based on the frozen status and the placement of modules to enable accurate pipeline stage balancing.
% We also incorporate the frozen status of modules and their placement within an MLLM to balance the pipeline stages.
% Unlike existing approaches that rely on the rule of thumb that backward pass execution time is approximately twice that of forward pass~\cite{pipedream-icml21,galvatron-vldb22}, we precisely measure backward pass times based on the frozen status of each module to enable accurate pipeline stage balancing.

\parabf{Workload balanced context parallelism for MLLMs (\S\ref{sec:context_parallelism}).}
\name introduces novel workload distribution algorithms that balance the computational cost of non-causal attention patterns at both inter-GPU and intra-GPU granularity.
At the inter-GPU granularity, we compute the computational cost of each token and distribute them so that GPUs have similar amount of workload as much as possible.
% During distribution, we break a previously held implicit assumption that all GPUs have the same number of tokens.
% Having different number of tokens per GPU may have more balanced distribution due to arbitrary attention patterns.
However, we notice that even if the GPUs have the same total amount of workload, the workload may not be evenly distributed to compute units within a GPU that leads to imbalance and performance degradation.
At the intra-GPU granularity, we shard the workload within each GPU to balance the computational cost across compute units.

We have implemented \name and conducted extensive evaluations on MLLMs of varying structures, modalities, and sizes.
Our evaluation results show that \name outperforms existing approaches by $2.26\times$ on average ($1.61\times$ -- $3.59\times$ across various model sizes) in training throughput.

To summarize, we make the following contributions:
\begin{denseitemize}
    \item We identify higher-order heterogeneity-borne challenges in MLLMs that affect the performance of distributed MLLM training.
    \item We design \name, a general-purpose distributed MLLM training framework that addresses those challenges.
    \item Our evaluation shows that \name surpasses existing approaches by $2.26\times$ on average in training throughput.
\end{denseitemize}

\section{Background and Motivation}
\label{sec:background}

We first introduce 4D-parallel distributed LLM training (\S\ref{sec:background-4d-parallelism}).
We then enumerate the unique characteristics of MLLMs that challenge existing LLM oriented training paradigms which motivates us to design \name (\S\ref{sec:background-characteristics-mllms}).

\subsection{4D Parallelism in LLM Training}
\label{sec:background-4d-parallelism}

Large-scale LLM training is a well-studied topic and leverages four parallelism dimensions -- tensor, pipeline, data, and context parallelism -- to achieve high training throughput~\cite{megatronlm-sc21,torchtitan-iclr25,llama3-arxiv24,deepspeed-sc20,megascale-nsdi24}.
Tensor and pipeline parallelism (TP and PP) partitions the model within each layer and across layers, respectively.
Data and context parallelism (DP and CP) partitions data; the former partitions a large batch of sequences into smaller minibatches, while the latter partitions each input sequence into segments.
In all parallelization dimensions, balancing the workload across GPUs is important to achieve high throughput.
In model partitioning, pipeline stages may have different amount of computation thus balancing them has been extensively studied~\cite{adapipe-asplos24,pipedream-icml21,galvatron-vldb22,alpa-osdi22,unity-osdi22}.
In data partitioning, the amount of workload can be imbalance across data parallel replicas and within each replica due to LLM's causal attention pattern~\cite{bytescale-arxiv25,ringattention-iclr24,wlbllm-osdi25,contextparallel-mlsys25}.

\subsection{Unique Characteristics of MLLMs}
\label{sec:background-characteristics-mllms}

% \begin{figure}[t]
% \centering
% \begin{subfigure}{\columnwidth}
%     \includegraphics[width=\textwidth]{figures/background_mllm_model_architecture.pdf}
%     \caption{An example of MLLM model architecture with two modality encoders and a LLM.}
%     \label{fig:background_mllm_model_architecture}
% \end{subfigure}

% \begin{subfigure}{\columnwidth}
%     \includegraphics[width=\textwidth]{figures/background_mllm_dataflow.pdf}
%     \caption{Dataflow of MLLM forward pass. Outputs from modality encoders are merged with text embedding and form an input sequence to the LLM.}
%     \label{fig:backgroun_mllm_dataflow}
% \end{subfigure}

% \caption{MLLM model architecture and dataflow.}
% \label{fig:background_mllm_characteristics}
% \end{figure}

MLLMs have unique characteristics that introduce new challenges to the existing 4D parallelism.

\parabf{Model and data heterogeneity.}
Unimodal LLMs usually contain repeated transformer layers with homogeneous structure, and unimodal text inputs go through the entire model.
Unlike unimodal LLMs, MLLMs have multiple modality encoders prior to the LLM with different structures.
The way of processing the input data is also different.
Modality encoders first process the modality input data that they are responsible for, and then project the output to the LLM.
The LLM then embeds the output of all modality encoders and text embedding, and computes the final output.

Model and data heterogeneity of MLLMs introduce imbalance in distributed training, which has been addressed by recent works.
For instance, DistMM~\cite{distmm-nsdi24}, DistTrain~\cite{disttrain-sigcomm25}, Optimus~\cite{optimus-atc25}, and DIP~\cite{dip-asplos26} disaggregate parallelism by applying different parallelization strategies to modalities to balance heterogeneous workload.

\begin{figure}[t]
    \centering
    \begin{subfigure}{\columnwidth}
        \includegraphics[width=\columnwidth]{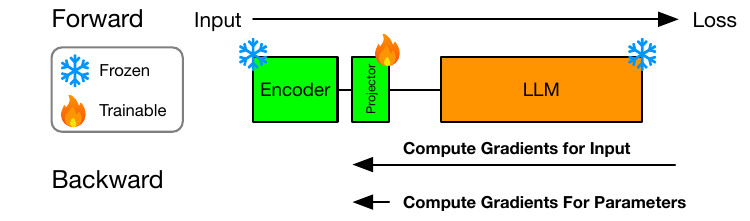}
        \caption{The frozen status and the location of layers alters the amount of gradient computation during the backward pass.}
        \label{fig:background_frozen_status_overview}
    \end{subfigure}
    \begin{subfigure}{\columnwidth}
        \centering
        \caption{Forward and backward pass execution time of the VLM with different combination of frozen status. Activation checkpointing is enabled~\cite{actrecomp-mlsys23}.}
        \label{tab:background_frozen_status_execution_time}
        \scriptsize
        \begin{tabular}{ccc|rr|rr}
            \toprule
            \multicolumn{3}{c|}{Trainable} & \multicolumn{2}{c|}{Encoder Time (ms)}             & \multicolumn{2}{c}{LLM Time (ms)}                   \\
            Enc & Proj & LLM               & \multicolumn{1}{c}{Fwd} & \multicolumn{1}{c|}{Bwd} & \multicolumn{1}{c}{Fwd} & \multicolumn{1}{c}{Bwd}  \\ 
            \midrule
            $\surd$   & $\surd$    & $\surd$                 & 44.43                   & 120.57                   & 140.39                  & 375.93                   \\
            $\times$   & $\surd$    & $\times$                 & 44.26                   & \textbf{0.50}                     & 138.33                  & \textbf{284.50}                   \\
            \bottomrule
        \end{tabular}
    \end{subfigure}
    \begin{subfigure}{\columnwidth}
        \includegraphics[width=\columnwidth]{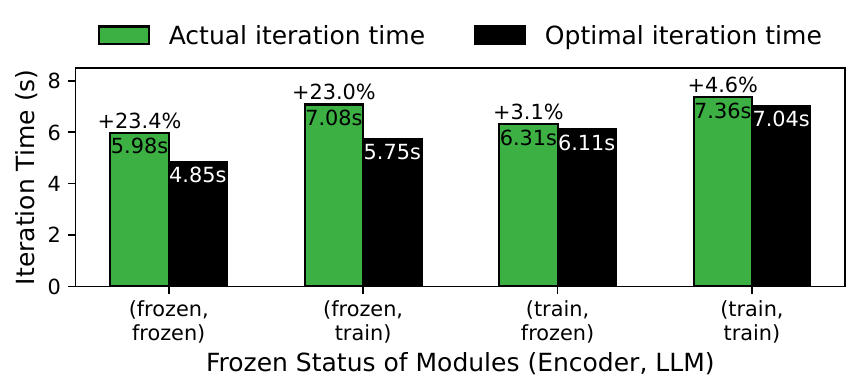}
        \caption{Execution time of the VLM with different combination of frozen status using pipeline parallelism on 4 NVIDIA A40 GPUs.
        The number of microbatch is 64.}
        \label{fig:background_frozen_status_imbalance}
    \end{subfigure}
    \caption{The impact of frozen status to the backward pass and the balance of pipeline stages. A VLM with Siglip (Encoder) and Llama-3.2 1b (LLM) is used.}
    \label{fig:background_frozen_status}
\end{figure}

\parabf{Using pretrained models with different frozen status.}
While unimodal LLMs are usually trained from scratch with randomly initialized parameters, MLLM training typically starts with a pretrained LLM and multiple unimodal encoders to exploit the representative capabilities of the pretrained models~\cite{llavanext-cvpr24,qwen2vl-arxiv24,internvl-cvpr24,llama3-arxiv24}.
Projectors between the modality encoders and the LLM are newly added and trained to align the embedding spaces of the modality encoders and the LLM.
Usually, the modality encoders and the LLM are frozen during MLLM training and only the projectors are trained~\cite{vlm-understanding-neurips24,survey-vlm-cvpr25}.

The frozen status of MLLM components significantly alters the amount of computation during the backward pass, as depicted in Figure~\ref{fig:background_frozen_status_overview} and Table~\ref{tab:background_frozen_status_execution_time}, invalidating the long-held rule of thumb that \textit{backward passes take roughly twice as long as forward passes}~\cite{pipedream-icml21}.
When pipeline parallelism is used, lack of considering the frozen status leads to execution time imbalance across pipeline stages.
Figure~\ref{fig:background_frozen_status_overview} illustrates the impact of frozen status to the balance of pipeline stages.
The model is partitioned to 4 pipeline stages to be balanced when all parameters are trainable, however, when the encoder is frozen, the balance between pipeline stages breaks, leading to slower execution.

\begin{figure}[t]
\centering
\begin{subfigure}{\columnwidth}
    \begin{minipage}[b]{0.4\textwidth}
        \includegraphics[width=\textwidth]{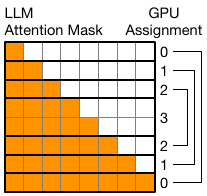}
    \end{minipage}%
    \hfill
    \begin{minipage}[b]{0.59\textwidth}
        \includegraphics[width=\textwidth]{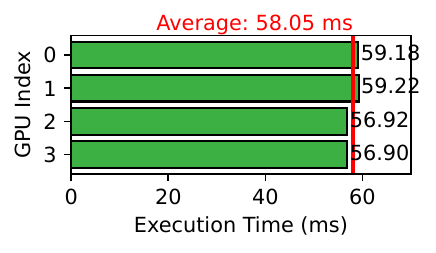}
    \end{minipage}
    \caption{Causal attention in LLM can easily be distributed evenly.}
    \label{fig:background_context_parallelism_llm}
\end{subfigure}
\begin{subfigure}{\columnwidth}
    \begin{minipage}[b]{0.4\textwidth}
        \includegraphics[width=\textwidth]{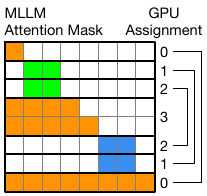}
    \end{minipage}%
    \hfill
    \begin{minipage}[b]{0.59\textwidth}
        \includegraphics[width=\textwidth]{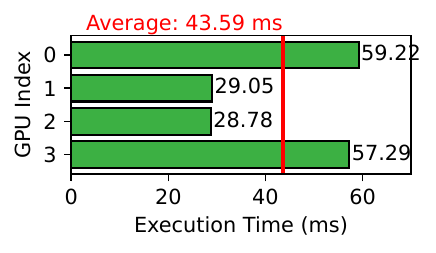}
    \end{minipage}
    \caption{Distributing MLLM attention patterns in the same way as in causal attention distribution is imbalanced.}
    \label{fig:background_context_parallelism_mllm}
\end{subfigure}

\caption{Balanced context parallelism optimized for LLMs. It is not applicable to MLLMs.}
\label{fig:background_context_parallelism}
\end{figure}

\parabf{Non-causal attention patterns break CP balance.}
In LLMs, a token attends to all preceding tokens and itself, forming a lower triangular matrix in the attention matrix, namely causal attention, as illustrated in Figure~\ref{fig:background_context_parallelism_llm}.
% Due to cross-modality interactions, however, MLLMs have more complex attention patterns as like in Figure~\ref{fig:background_context_parallelism_mllm}~\cite{flexattention-mlsys25,flashmask-iclr25}.
Most existing context parallelism works exploit the causal attention pattern~\cite{wlbllm-osdi25,contextparallel-mlsys25}.
They partition the sequence into $2 \times \text{cp\_size}$, where $\text{cp\_size}$ is the number of ranks in context parallelism dimension, and the $i$-th rank gets $i$-th and $(2 \times \text{cp\_size} - 1 - i)$-th chunks.
This distribution guarantees balanced workload in causal attention.

However, MLLMs have non-causal attention patterns due to cross-modality interactions, as shown in Figure~\ref{fig:background_context_parallelism_mllm}, where the same distribution is imbalanced.
Moreover, the attention pattern is variable depending on the input data, different from LLMs that have fixed causal attention pattern.
Statically distributing the workload assuming fixed attention pattern is not applicable to MLLMs.

DCP~\cite{dcp-sosp25} is the first work that considers dynamic non-causal context parallelism.
However, its design is based on ring context parallelism~\cite{ringattention-iclr24}, which has proven to be inefficient in large scale training and replaced by All-Gather based context parallelism~\cite{loongtrain-corr24, llama3training-isca25}.

\section{Multimodality-Aware Parallelization}
\label{sec:model_parallelization}

Based on the observation in Section~\ref{sec:background}, we introduce \name's MLLM-specific parallelization strategies.
We first introduce frozen status-aware pipeline parallelism (\S\ref{sec:pipeline_parallelism}) and then workload-balanced context parallelism (\S\ref{sec:context_parallelism}).

\subsection{Frozen Status-Aware Pipeline Parallelism}
\label{sec:pipeline_parallelism}

% TODO: generate a figure

\name's frozen status-aware pipeline parallelism partitions the model into pipeline stages where the sum of one forward execution time and one backward execution time (1F+1B) is balanced across stages considering the frozen status of the layers.

However, simply considering the frozen status and adopting all-or-nothing approach -- add backward pass computation time if trainable, otherwise skip -- is not correct either.
Even if a layer is frozen, it may still need to backpropagate gradients to the trainable parameters ahead of it.
Backward pass computation of a layer $L_l$ consists of two parts: gradient computation for parameters $B_{wl}$ and gradient computation for data $B_{dl}$~\cite{zerobubblepp-iclr24}:
\begin{align}
  B_l = B_{wl} + B_{dl}
\end{align}
If trainable parameters are located ahead of a frozen layer, the frozen layer, while it can skip the gradient computation for its parameters (\ie, $B_{w} = 0$), needs to \textit{backpropagate input gradients} to the trainable parameters (\ie, $B_{d} \neq 0$), so that the trainable parameters can update themselves.
Then, the backward pass execution time $B_l$ can be computed as:
\begin{equation}
  \begin{split}
  B_l = & (B_{wl} \text{ if } f(L_l) \text{ is False else } 0) \\
  + & (B_{dl} \text{ if } p(L_l) \text{ is True else } 0)
  \end{split}
\end{equation}
where $f(L_l)$ returns the frozen status of the $l$-th layer $L_l$ and $p(L_l)$ returns whether the $l$-th layer $L_l$ has trainable parameters ahead of it.
$p(L_l)$ exhibits forward propagation; once it is set to True at some layer that is trainable, all the layers after it need to have $p(L)$ True to backpropagate gradients.
Thus, $p(L_l)$ can be computed as:
\begin{equation}
  \begin{split}
  p(L_l) = & \begin{cases}
    \text{True} & \text{if } p(L_{l-1}) \text{ or not } f(L_l) \\
    \text{False} & \text{otherwise}
  \end{cases} \\
  p(L_0) = & \text{ not } f(L_0)
  \end{split}
\end{equation}

Checking trainable parameters ahead of the layer should also be done across modalities.
If some parameters in the modality encoders are trainable, all layers after it in the same modality as well as the LLM need to backpropagate gradients, since the LLM sits after the modality encoders.

The per-layer cost $T_l = F_l + B_l$ derived from these rules is then used to partition the model into $K$ balanced pipeline stages, minimizing the bottleneck stage cost.
Any off-the-shelf partitioning algorithms -- such as dynamic programming~\cite{alpa-osdi22} or divide and conquer~\cite{oobleck-sosp23} -- can be used to partition the model into pipeline stages, as long as it accepts the per-layer cost $T_l$ as input.

\subsection{Token Workload-Balanced Context Parallelism}
\label{sec:context_parallelism}

\begin{figure}[t]
  \centering
  \begin{subfigure}[t]{\columnwidth}
    \includegraphics[width=\columnwidth]{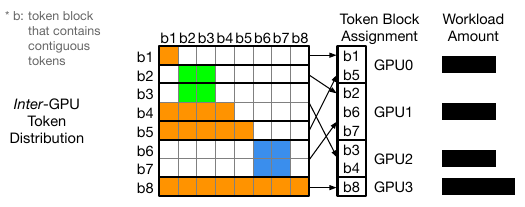}
    \caption{Inter-GPU workload balancing. Tokens are distributed across GPUs.}
    \label{fig:context_parallelism_inter}
  \end{subfigure}
  \begin{subfigure}[t]{\columnwidth}
    \includegraphics[width=\columnwidth]{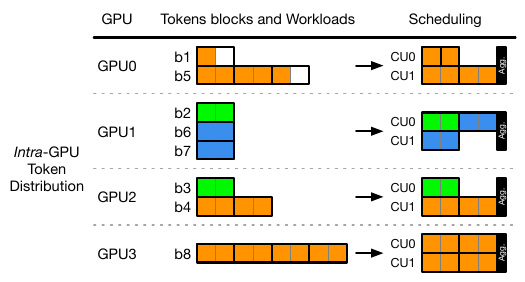}
    \caption{Intra-GPU workload balancing. Each token has at most 8 blocks to compute. These blocks are partitioned into subblocks of up to 2 blocks each, creating at most 4 subblocks per token that are scheduled to CUs.}
    \label{fig:context_parallelism_intra}
  \end{subfigure}
  \caption{Two-step (inter-GPU and intra-GPU) workload balanced context parallelism.}
  \label{fig:context_parallelism}
\end{figure}

In multimodal context parallelism, many non-causal attention masks can be generated~\cite{blip2-icml23,gemma3-arxiv25,flashmask-iclr25,flexattention-mlsys25}, which the existing token distribution for LLMs fails to balance.
We observe that to achieve genuine workload-balanced context parallelism, workload distribution across and within GPUs should be considered simultaneously.
We call them \textit{inter-GPU} and \textit{intra-GPU} workload balancing and discuss in Section~\ref{sec:inter_gpu_balance} and Section~\ref{sec:intra_gpu_balance}, respectively.

\subsubsection{Inter-GPU Workload Balancing}
\label{sec:inter_gpu_balance}

Inter-GPU workload imbalance indicates that the amount of workloads distributed to each GPU is not balanced.
This is because modern attention implementations introduce variations in the amount of computation per token~\cite{flashattn-nips22,flashattn2-iclr24,flexattention-mlsys25}.
They partition tokens into blocks and skip block computations for efficiency if the corresponding block is completely masked-out.
The amount of workloads to compute attention output per query token block can be computed by counting the number of colored blocks \textit{rowwise}.
In Figure~\ref{fig:context_parallelism_inter}, for example, the workloads of 8 blocks are 1, 2, 2, 4, 5, 2, 2, 8, respectively, which are varied and irregular.
We therefore propose a new method of distributing the tokens across GPUs based on the amount of computations, which we call \textit{inter-GPU workload-balanced distribution}.

\parabf{Problem formulation.}
We first formulate the problem as an integer linear programming (ILP) problem as follows:
\begin{mini}[2]
{x, C}
{C}{}{}
% Constraints
\addConstraint{\sum_{g=1}^{G} x_{i,g}}{= 1,}{i = 1, \dots, T}
\addConstraint{\sum_{i=1}^{T} W_i \cdot x_{i,g}}{\leq C, \quad}{g = 1, \dots, G}
\addConstraint{x_{i,g}}{\in \{0, 1\}}
\end{mini}
Here, $x_{i,g}$ is a binary decision variable that indicates whether token $i$ is assigned to $g$-th GPU over $G$ GPUs. 
$W_i$ represents the workload of $i$-th token $x_i$, which can be computed by row-wise sum of unmasked part of the attention mask that needs computation.
The linear programming balances workload by minimizing the completion time $C$, which is the maximum workload assigned to any GPU.

\begin{algorithm}[t]
  \caption{Token workload-balanced context parallelism algorithm.}
  \label{alg:context_parallelism}
  \begin{algorithmic}
    \STATE \textbf{Input: } Tokens $T$, block size $N_B$, \# GPUs $G$, and attention mask $A$
    \STATE \textbf{Output: } Token assignment to GPUs $X_0, X_1, \dots, X_{G-1}$

    \STATE $B \leftarrow$ partition $T$ into blocks of size $N_B$
    \FOR{$b \in B$}
    \STATE $W_b \leftarrow$ \# blocks to compute in attention $A$ for $b$
    \ENDFOR

    \STATE $L \leftarrow$ \texttt{minheap()}, $X \leftarrow$ \texttt{dict()}
    \FOR{$g \in 0, \dots, G-1$}
      \STATE \texttt{L.heappush($g, 0$)} \label{line:initialize_gpu_workload}
      \STATE \texttt{$X[g]$ = list()}
    \ENDFOR

    \FOR{$b, W_b \in B, W$}
      \STATE $g, W[g] \leftarrow$ \texttt{L.heappop()}
      \STATE \texttt{$X[g]$.append($b$)}
      \STATE \texttt{L.heappush($g, W[g] + W_b$)}
    \ENDFOR

    \STATE \textbf{return} $X$
  \end{algorithmic}
\end{algorithm}

\parabf{Weighted makespan minimization.}
For a long sequence, the ILP problem is intractable in real-time during training, thus we adopt the greedy Longest-Processing-Time-First (LPT) algorithm to assign tokens to GPUs in a context parallelism group for fast and efficient distribution~\cite{graham-scheduling-siam69}.
Algorithm~\ref{alg:context_parallelism} shows an adapted LPT algorithm that considers the characteristics of parallel accelerators that compute with a large amount of data.
We first partition the tokens into blocks of size $N_B$ (\eg, 128).
For each token block, we count the number of blocks to compute to measure the workload of the token.
If the corresponding attention mask block is full of zeros, the block is skipped.
We then use the LPT algorithm to assign the token block to the GPU with the least amount of workload assigned so far.

The longest processing time in the worst case has proven to be $\sum_{i=0}^{T-1}\frac{t_i}{G} + t_{\text{max}}$, where $i$-th token's amount of attention computation is $t_i$, total number of tokens $T$, and the number of GPUs $G$~\cite{graham-scheduling-siam69}.
As $T$ increases, $\sum\frac{t_i}{G}$ dominates the processing time, and it is getting closer to the perfectly balanced distribution.
It requires $O(GTlogT)$ time complexity, where $TlogT$ is consumed by sorting the tokens in descending order of their workloads. 

\subsubsection{Intra-GPU Workload Balancing}
\label{sec:intra_gpu_balance}
Even with inter-GPU workload balanced distribution, which evenly distributes the \textit{total amount of computation} across GPUs, architectural characteristics of GPUs and implementation of attention can still lead to imbalanced execution when the jobs are dispatched to compute units (CUs).

Revisiting modern attention implementations~\cite{onlinesoftmax-arxiv18,memory-efficient-attention-arxiv21,flashattn-nips22,flashattn2-iclr24,flexattention-mlsys25}, they are designed to avoid unnecessary memory accesses as much as possible.
CUs use online softmax algorithm and compute the final attention output of a single query token block by keeping the intermediate output in the cache and iterating over the entire key and value blocks in a single kernel.
This minimizes the number of memory accesses by not writing intermediate variables to global memory.

However, assigning attention computation of a block \textit{as a whole} to a CU introduces imbalanced amount of workload across CUs.
In Figure~\ref{fig:context_parallelism_intra}, for example, b1 and b5 assigned to GPU0 are executed in parallel on CU0 and CU1, respectively.
While the amount of computation of b1 (1 block) and b5 (5 blocks) are extremely different, computing b5 cannot be parallelized across CUs; thus, CU0 has to wait for CU1 to finish before proceeding to the next kernel execution.

We observe that the idea of blockwise parallel attention, which was originally designed to parallelize attention across multiple accelerators, can also be used to balance the workload across CUs in a single GPU~\cite{blockwise-neurip23}.
We adopt it for intra-GPU workload balancing, where the attention computation of a single set of query tokens is split into multiple subblocks and scheduled in parallel.
Figure~\ref{fig:context_parallelism_intra}, for example, partitions attention computations to subblocks of size 2.
Each kernel computes partial attention output for a single subblock and writes it to local memory. Then, we launch an additional aggregation kernel that gathers the local attention outputs and computes the final attention output for all query blocks.
Unlike the original attention computation, which writes the final attention output to memory, our output is local attention output per subblock that needs to be aggregated.
The size of blocks (\eg, 2 subblocks per block in Figure~\ref{fig:context_parallelism_intra}) affects the performance; with smaller subblocks, workloads are more balanced, but more local outputs should be written to the global memory, and then read again for aggregation.
Large subblocks, on the other hand, have less overhead of aggregation but more imbalance.
We empirically find that using 16 -- 32 subblocks per block achieves good performance.

\section{Implementation}
\label{sec:implementation}

% \begin{figure}[t]
%     \centering
%     \includegraphics[width=\columnwidth]{figures/cornstarch_design.pdf}
%     \caption{\name architectural overview.}
%     \label{fig:cornstarch_overview}
% \end{figure}

% Figure~\ref{fig:cornstarch_overview} shows the architectural overview of \name.
% Given a MLLM, \name first partitions the model considering their amount of workload.
% Once model parallelization is done, \name starts training the model by feeding the data to the model after applying the data parallelization.
% A batch is partitioned into multiple microbatches, and sequences in each microbatch are again sharded into multiple sets of tokens, which have balanced workload.

% This section explains details about \name internal implementation.
\name is implemented in around 26k new Python SLOC on top of PyTorch 2.6.0~\cite{pytorch-neurips19}, HuggingFace Transformers 4.51.0~\cite{transformers-acl20}, and Colossal-AI 0.4.6~\cite{colossalai-icpp23}.
\name's model partitioning, scheduling, execution, communication, and checkpointing are implemented upon Colossal-AI interface.
\name supports various model families and model sizes so that users can train mode than 10,000 different combinations of MLLMs.
% All unimodal models in the supported model families available in the HuggingFace hub can be used in creating an MLLM~\cite{hfhub-web}.
See Appendix~\ref{sec:apdx-model-list} for the list of supported models.

\begin{figure}[t]
    \centering
    \includegraphics[width=\columnwidth]{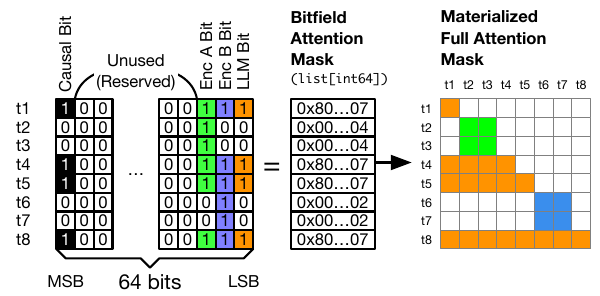}
    \caption{Bitfield attention mask representation.}
    \label{fig:bitfield_attention}
\end{figure}

\subsection{Implementation of Pipeline Stage Partitioning}
\label{sec:impl_pipeline_parallelism}

\name assigns each layer a per-layer cost from Section~\ref{sec:pipeline_parallelism} and partitions the model into $K$ contiguous pipeline stages.
The partitioner minimizes the bottleneck stage cost -- the 1F+1B time of the slowest stage -- which is the standard objective for pipeline stage balancing~\cite{alpa-osdi22,oobleck-sosp23}.
\name solves this contiguous partitioning problem with dynamic programming~\cite{alpa-osdi22}; the search space is small in practice (hundreds of layers and up to tens of stages), so partitioning completes in negligible time relative to training.
Let $w(j,i) = \sum_{\ell=j}^{i-1} T_\ell$ be the cost of the contiguous segment from layer $j$ to $i{-}1$.
$X[i][k]$ is the optimal value when the first $i$ layers are partitioned into $k{+}1$ stages:
\begin{align}
  X[i][0] &= w(0,i) \label{eq:dp_base} \\
  X[i][k] &= \min_{j \,\in\, [k,\, i)} \max\!\bigl(X[j][k-1],\; w(j,i)\bigr) (k > 0) \label{eq:dp_recurrence}
\end{align}
The recurrence tries every split point $j$: the bottleneck is the larger of the optimal bottleneck for the prefix of length $j$ with $k$ stages and the cost $w(j,i)$ of the last stage.
% \name fills $X$ bottom-up and, for each $(i,k)$, stores the argmin $j$ from \eqref{eq:dp_recurrence}.
% The optimal bottleneck cost is $X[N][K{-}1]$.
\name records the argmin $j$ at each subproblem; the optimal bottleneck cost is $X[N][K{-}1]$.
To obtain the partition, \name backtracks from $(N, K{-}1)$: the stored $j$ marks where the last stage begins, so layers $j$ through $N{-}1$ form the final stage, and the algorithm recurses on the prefix subproblem $(j, K{-}2)$.
Repeating until $k=0$ yields all $K$ contiguous stage boundaries.

\subsection{Implementation of MLLM Attention}
\label{sec:impl_mllm_attention}

\name implements \textit{bitfield attention} in Triton~\cite{triton-mapl19,triton-github21} for high performance non-causal attention execution.
Naive attention implementation computes attention scores for all tokens, and then applies the attention mask as a whole to the attention scores.
This is inefficient as it requires high memory bandwidth and is not parallelizable.
Bitfield attention mask is a sparse representation of the attention mask to represent multimodal interactions into attention patterns efficiently.
Full attention mask is a very large 4D tensor (batch $\times$ \# heads $\times$ sequence length $\times$ sequence length), which needs too much memory for long sequences.
Bitfield attention mask is a 2D 64-bit integer tensor (batch $\times$ sequence length), where each bit represents which modalities the token at that position needs to attend to.
Figure~\ref{fig:bitfield_attention} shows an example of a bitfield attention mask.
We assign bits from the least significant bit (LSB) to the most significant bit (MSB) to the modality encoders and the LLM.
The LSB (1st index) is assigned to the LLM, and 2nd and 3rd bits are assigned to the modality encoders A and B, respectively, for example.
The most significant bit (64th index) is reserved for causal bit; when this bit is set to 1, the token attends to all of its previous tokens.
For example, tokens \texttt{t2, t3} are tokens from the encoder A, thus have 2nd LSB set to 1. As it does not have causal bit, it attends to the other tokens only from the encoder A.
\texttt{t4, t5}, however, are text tokens with causal bit and all modality bits set to 1.
Thus it can attend to all modality tokens, but only previous tokens.
\name bitfield attention implementation is compatible with context parallelism (\S\ref{sec:impl_context_parallelism}), while standard FlashAttention does not natively support.
% \name bitfield attention implementation is compatible with context parallelism (\S\ref{sec:context_parallelism} and \S\ref{sec:impl_context_parallelism}) and can distribute tokens and attention patterns while other implementations for complex attention patterns do not support context parallelism~\cite{flashmask-iclr25,flexattention-mlsys25}.

\subsection{Implementation of Context Parallelism}
\label{sec:impl_context_parallelism}
There are various ways of implementing context parallelism: all-to-all~\cite{ulysses-arxiv23}, ring attention~\cite{ringattention-iclr24,dcp-sosp25}, and All-Gather based~\cite{llama3-arxiv24,llama3training-isca25}, \etc\xspace
% All-to-all has no imbalance problem as it converts parallelization dimension from the token space to the head space and each rank computes exactly the same shape of attention.
% However, it has a high communication overhead that cannot be overlapped with computation, and it has limited scalability because its maximum parallelization degree is constrained by the number of heads~\cite{loongtrain-corr24}.
% \mosharaf{Confusing sentence.}
\name implements the SOTA All-Gather based context parallelism implementation.
This implementation gathers all keys and values of all tokens and compute row-wise attention for local queries.
Overlapping communication and computation is done in the head dimension; while GPUs compute attention for one or a few heads, it transfers keys and values for the next head(s).
This simplifies Algorithm~\ref{alg:context_parallelism} in computing per-token workload.
If we adopt P2P ring attention, it would have been more complicated to compute per-token workload as it requires to recompute the amount of workloads every round.

\section{Evaluation}
\label{sec:evaluation}

In this section, we evaluate \name and show its effectiveness in training MLLMs.
Our key results are:
\begin{denseitemize}
    \item \name achieves $2.26\times$ higher end-to-end training throughput on average for MLLM training (\S\ref{sec:eval-e2e}).
    
    \item Frozen status-aware pipeline parallelism partitions MLLMs more effectively by considering the frozen status and provides up to 2.46$\times$ faster iteration time in MLLMs (\S\ref{sec:eval-pipeline-parallelism}).
    
    \item Workload-balanced context parallelism distributes tokens more evenly across GPUs and within a single GPU, which improves the performance of attention execution by up to 1.18$\times$ (\S\ref{sec:eval-context-parallelism}).
\end{denseitemize}

\subsection{Experimental Setup}
\label{sec:experimental-setup}
\parabf{Testbed.}
We run our evaluation workloads in a GPU cluster with 6 nodes, each with four NVIDIA A40-48GB GPUs and a NVIDIA Mellanox ConnectX-6 200Gbps Infiniband adaptor (total 24 GPUs). The four GPUs in a node are connected in pairs using NVLink and connected to the node via PCIe 4.0.

\parabf{Baselines.}
We set the baselines as follows:

\begin{denseenum}
    \item \textit{FSDP}: FSDP is widely used in distributed MLLM training thanks to its ease of use~\cite{internvl-cvpr24,cogvlm-arxiv24,llava-neurips23,llavanext-cvpr24}.
    It shards parameters and distributes them across all GPUs to reduce memory footprint.
    Parameters are temporarily gathered for computation and then sharded again. We use FSDP2, which offers higher performance~\cite{torchtitan-iclr25}.
    \item \textit{Megatron*}: Megatron-LM extends LLM pipeline parallelism to MLLMs by adding a vision encoder as the first pipeline stage~\cite{megatronlm-github}.
    We chose Megatron-LM as a representative of the existing LLM-optimized 4D parallelization.
\end{denseenum}

\begin{table}[tb]
\centering
\small
\caption{Modality (LLM, vision, and audio) configurations.}
\label{tab:model-selection}
\begin{tabular}{cc|cc|c} 
    \toprule
    \begin{tabular}[c]{@{}c@{}}Model\\Arch.\end{tabular}                   & \begin{tabular}[c]{@{}c@{}}Model\\Size\end{tabular} & \# Layers & \begin{tabular}[c]{@{}c@{}}Hidden\\Size\end{tabular} & \# Params  \\ 
    \midrule
    \multirow{3}{*}{\begin{tabular}[c]{@{}c@{}}Llama-3 \\(LLM)\end{tabular}} & Small                                               & 16        & 2048                                                 & 1b         \\
                                                                            & Medium                                              & 32        & 4096                                                 & 8b         \\
                                                                            & Large                                               & 64        & 5120                                                 & 32b        \\ 
    \midrule
    \multirow{3}{*}{\begin{tabular}[c]{@{}c@{}}Qwen2\\Vision\end{tabular}} & Small                                               & 32        & 1280                                                 & 0.6b       \\
                                                                            & Medium                                              & 48        & 2560                                                 & 3.9b       \\
                                                                            & Large                                               & 64        & 3840                                                 & 11.6b       \\ 
    \midrule
    \multirow{3}{*}{\begin{tabular}[c]{@{}c@{}}Phi4\\Audio\end{tabular}}   & Small                                               & 24        & 1024                                                 & 0.5b       \\
                                                                            & Medium                                              & 32        & 3072                                                 & 3.4b       \\
                                                                            & Large                                               & 48        & 5120                                                 & 12.4b       \\
    \bottomrule
    \end{tabular}
\end{table}

\parabf{Training data.}
We use a synthetic dataset for evaluation.
Each sample consists of 1k text tokens, a 1280x720 image, and a 2-minute audio clip.
Image tokens and audio tokens are projected into the text embedding space after being processed by the corresponding modality encoder.
We use a global batch size of 48. FSDP uses a mini-batch size of 2, while microbatch size 4 is used in Megatron* and \name with pipeline parallelism.
The attention patterns are described in Appendix~\ref{sec:apdx-attention-patterns}.

\parabf{Model configurations.}
We evaluate various MLLM configurations created by combining two modality encoders (vision and audio) and an LLM, each selected from the sizes listed in Table~\ref{tab:model-selection}.
The modality encoders are merged into a single module which processes both modalities.
We freeze the merged modality encoder and the LLM and only train the projector modules.
An MLLM configuration is denoted by suffixes representing the sizes (S, M, L) of its vision encoder, audio encoder, and LLM, respectively (\eg, MLLM-SML combines a small vision encoder, a medium audio encoder, and a large LLM).

\begin{figure}[t]
    \centering
    \begin{subfigure}[rt]{\columnwidth}
        \includegraphics[width=\columnwidth]{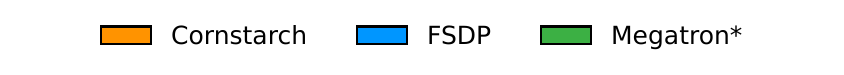}
    \end{subfigure}
    \begin{subfigure}[t]{\columnwidth}
        \includegraphics[width=\columnwidth]{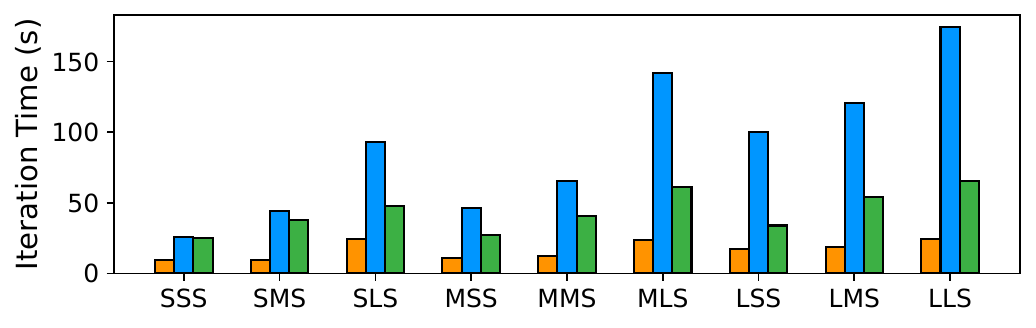}
        \caption{MLLMs with small LLM.}
        \label{fig:e2e-mllm-s}
    \end{subfigure}
    \begin{subfigure}[t]{\columnwidth}
        \includegraphics[width=\columnwidth]{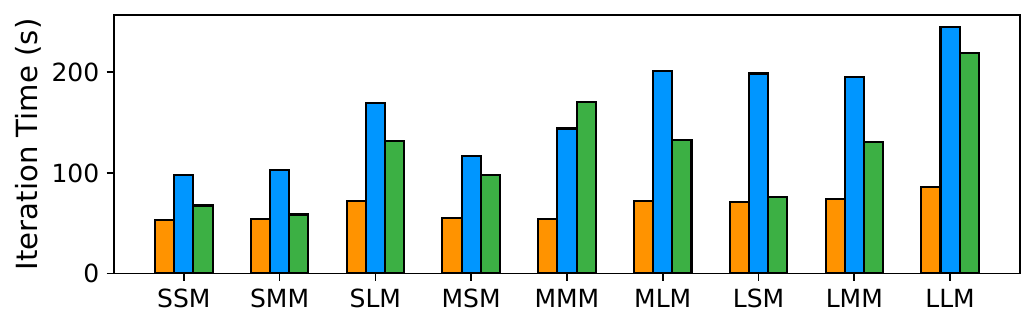}
        \caption{MLLMs with medium LLM.}
        \label{fig:e2e-mllm-m}
    \end{subfigure}
    \begin{subfigure}[t]{\columnwidth}
        \includegraphics[width=\columnwidth]{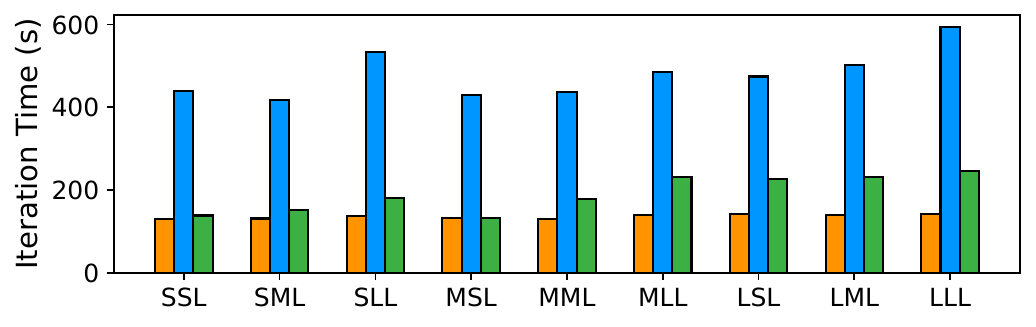}
        \caption{MLLMs with large LLM.}
        \label{fig:e2e-mllm-l}
    \end{subfigure}
    \caption{End-to-end performance comparison of \name and baselines with various model configurations.}
    \label{fig:e2e}
\end{figure}

\subsection{End-to-End Performance}
\label{sec:eval-e2e}

We first evaluate \name against the baselines in terms of end-to-end training iteration time and show the results in Figure~\ref{fig:e2e}.
When models are small enough (\eg,~MLLM-SSS in Figure~\ref{fig:e2e-mllm-s}), FSDP shows reasonably good performance.
However, as model size increases, FSDP's performance drops significantly due to intensive communication overhead.
For Megatron*, its current limitations cause a large imbalance in pipeline stages due to the lack of frozen status awareness.
This is observed especially well when the modality encoders are relatively larger than LLMs (\eg, MLLM-LLS in Figure~\ref{fig:e2e-mllm-s} or MLLM-LLM in Figure~\ref{fig:e2e-mllm-m}).
% While GraphPipe* can execute multiple modality encoders in parallel, its lack of context parallelism support leads to worse performance than Megatron* in cases where executing modality encoders in parallel does not help.
% For Megatron*, its current limitation that can have at most one pipeline stage for modality encoders causes a large imbalance in pipeline stages, especially when encoders are relatively large (\eg,~MLLM-LLS in Figure~\ref{fig:e2e-mllm-s}).
% GraphPipe*, which is based on graph pipeline parallelism, shows better performance than Megatron* but only with the small LLM.
% The large LLM needs more pipeline stages to balance the execution time.
% However, modality encoders need dedicated GPUs to run in parallel, which leaves fewer GPUs for the LLM.
% As a result, pipeline stages for the LLM become stragglers, leading to performance degradation.

\name shows the best performance across all model configurations.
It chooses better modality parallelism, allocates pipeline stages to the encoders and the LLM based on their frozen status and their placement, and balances the attention computation on the fly.
We discuss the performance of \name in detail in the subsequent sections.
Overall, \name outperforms the baselines 2.26$\times$ on average (3.36$\times$ vs FSDP and 1.62$\times$ vs Megatron*).

\subsection{Impact of Frozen Status-Aware Pipeline Parallelism}
\label{sec:eval-pipeline-parallelism}

\begin{table}[tb]
\centering
\tiny
\caption{Model forward and backward execution time breakdown parallelized with and without frozen status awareness.}
\label{tab:frozen-aware-pipeline-parallelism}
\begin{tabular}{cc|rr|rr|rc} 
\toprule
\multirow{2}{*}{\begin{tabular}[c]{@{}c@{}}Model\end{tabular}} & \multirow{2}{*}{\begin{tabular}[c]{@{}c@{}}Frozen\\Aware\end{tabular}} & \multicolumn{2}{c|}{\begin{tabular}[c]{@{}c@{}}Per-Stage\\Fwd (ms)\end{tabular}} & \multicolumn{2}{c|}{\begin{tabular}[c]{@{}c@{}}Per-Stage\\Bwd (ms)\end{tabular}} & \multicolumn{1}{c}{\multirow{2}{*}{\begin{tabular}[c]{@{}c@{}}Iter.\\Time\\(s)\end{tabular}}} & \multirow{2}{*}{\begin{tabular}[c]{@{}c@{}}Impr.\\($\times$)\end{tabular}}  \\ 
\cmidrule(l){3-6}
                                                                        &                                                                        & \multicolumn{1}{c}{Enc} & \multicolumn{1}{c|}{LLM}                               & \multicolumn{1}{c}{~Enc~} & \multicolumn{1}{c|}{LLM}                             & \multicolumn{1}{c}{}                                                                             &                                                                            \\ 
\midrule
\multirow{2}{*}{SSS}                                                 & $\surd$                                                  & 301.61                  & 149.40                                                 & 1.04                      & 518.43                                               & 21.81                                                                                            & 1.18x                                                                      \\
                                                                        & $\times$                                                  & 207.13                  & 296.25                                                 & 0.86                      & 1032.63                                              & 25.67                                                                                            & -                                                                          \\ 
\midrule
\multirow{2}{*}{MMS}                                                 & $\surd$                                                  & 903.86                  & 102.61                                                 & 2.72                      & 346.54                                               & 40.34                                                                                            & 1.02x                                                                      \\
                                                                        & $\times$                                                  & 635.56                  & 297.62                                                 & 1.19                      & 1032.12                                              & 41.21                                                                                            & -                                                                          \\ 
\midrule
\multirow{2}{*}{LLS}                                                 & $\surd$                                                  & 1240.50                 & 298.37                                                 & 1.41                      & 1029.98                                              & 66.14                                                                                            & 1.00x                                                                      \\
                                                                        & $\times$                                                  & 1259.13                 & 297.92                                                 & 1.15                      & 1030.14                                              & 66.20                                                                                            & -                                                                          \\ 
\midrule
\multirow{2}{*}{SSM}                                                 & $\surd$                                                  & 464.98                  & 331.43                                                 & 3.53                      & 3017.01                                              & 66.56                                                                                            & 1.11x                                                                      \\
                                                                        & $\times$                                                  & 388.33                  & 388.73                                                 & 2.30                      & 4030.74                                              & 73.94                                                                                            & -                                                                          \\ 
\midrule
\multirow{2}{*}{MMM}                                                 & $\surd$                                                  & 2330.90                 & 273.89                                                 & 2.09                      & 2418.27                                              & 70.37                                                                                            & \textbf{2.46x}                                                             \\
                                                                        & $\times$                                                  & 712.72                  & 1159.76                                                & 1.60                      & 12113.67                                             & 173.01                                                                                           & -                                                                          \\ 
\midrule
\multirow{2}{*}{LLM}                                                 & $\surd$                                                  & 2199.91                 & 376.11                                                 & 4.15                      & 4023.22                                              & 87.43                                                                                            & 2.03x                                                                      \\
                                                                        & $\times$                                                  & 1403.97                 & 1161.76                                                & 1.56                      & 12109.73                                             & 177.39                                                                                           & -                                                                          \\ 
\midrule
\multirow{2}{*}{SSL}                                                 & $\surd$                                                  & 773.10                  & 741.17                                                 & 3.55                      & 6309.11                                              & 138.45                                                                                           & 1.00x                                                                      \\
                                                                        & $\times$                                                  & 774.19                  & 708.62                                                 & 3.84                      & 6306.18                                              & 137.62                                                                                           & -                                                                          \\ 
\midrule
\multirow{2}{*}{MML}                                                 & $\surd$                                                  & 2280.58                 & 705.73                                                 & 3.11                      & 6311.23                                              & 138.97                                                                                           & 1.30x                                                                      \\
                                                                        & $\times$                                                  & 1015.60                 & 1154.85                                                & 2.78                      & 10546.34                                             & 180.28                                                                                           & -                                                                          \\ 
\midrule
\multirow{2}{*}{LLL}                                                 & $\surd$                                                  & 5316.08                 & 736.05                                                 & 4.00                      & 6315.46                                              & 143.76                                                                                           & 1.72x                                                                      \\
                                                                        & $\times$                                                  & 1597.36                 & 1686.28                                                & 3.06                      & 15878.58                                             & 247.79                                                                                           & -                                                                          \\
\bottomrule
\end{tabular}
\end{table}

We parallelize the models with frozen status-aware pipeline parallelism and compare the performance with the same models but parallelized without frozen status awareness.
Table~\ref{tab:frozen-aware-pipeline-parallelism} presents the results.
For brevity, we only show a few model configurations with encoders being colocated.

Without frozen status-awareness, partitioning is done based on the assumption of all parameters being trainable, which tries to minimize variance of forward time across pipeline stages.
For example, MLLM-LLL, the frozen status-unaware partitioning partitions the modality encoders and the LLM to have similar forward execution time ($\sim$ 1600ms).
However, gradient computations for the frozen encoders and the LLM are skipped, their backward execution time is significantly different (3.06ms and 15878.58ms), breaking the balance between pipeline stages.

With frozen status-awareness, the partitioning is balanced based on the forward execution time plus the backward execution time (5320.08ms and 7051.51ms, respectively), decreasing pipeline bubbles.
We also observe a few exceptions: MLLM-LLS and MLLM-SSL.
These are the most extreme cases in model size distinction, thus even assigning maximum number of pipeline stages to the large module is not enough to balance the pipeline stages.
In other cases, frozen status-aware pipeline parallelism assigns workloads more evenly across pipeline stages, which improves the overall performance by up to 2.46$\times$.

\subsection{Impact of Workload-Balanced Context Parallelism}
\label{sec:eval-context-parallelism}

\begin{table}[tb]
\scriptsize
\centering
\caption{Execution time of a single attention layer and entire LLM with 64k sequence length using various context parallelization policies.}
\label{tab:context_parallel}
\begin{tabular}{cc|rrrr} 
    \toprule
    \multicolumn{2}{c|}{\begin{tabular}[c]{@{}c@{}}Time (ms)\\(Impr. ($\times$))\end{tabular}} & \multicolumn{1}{c}{\begin{tabular}[c]{@{}c@{}}Causal\\CP\end{tabular}} & \multicolumn{1}{c}{\begin{tabular}[c]{@{}c@{}}Inter-GPU\\Balance\\Only\end{tabular}} & \multicolumn{1}{c}{\begin{tabular}[c]{@{}c@{}}Intra-GPU\\Balance\\Only\end{tabular}} & \multicolumn{1}{c}{\name}                                     \\ 
    \midrule
    \multirow{2}{*}{LLM-S} & Attn                                                   & \begin{tabular}[c]{@{}r@{}}243.44\\(-)\end{tabular}                    & \begin{tabular}[c]{@{}r@{}}255.59\\(0.95x)\end{tabular}                              & \begin{tabular}[c]{@{}r@{}}225.73\\(1.08x)\end{tabular}                              & \begin{tabular}[c]{@{}r@{}}204.95\\\textbf{(1.19x)}\end{tabular}  \\
                           & Model                                                  & \begin{tabular}[c]{@{}r@{}}5541.25\\(-)\end{tabular}                   & \begin{tabular}[c]{@{}r@{}}5665.77\\(0.98x)\end{tabular}                             & \begin{tabular}[c]{@{}r@{}}5250.40\\(1.06x)\end{tabular}                             & \begin{tabular}[c]{@{}r@{}}4856.60\\(1.14x)\end{tabular}          \\ 
    \midrule
    \multirow{2}{*}{LLM-M} & Attn                                                   & \begin{tabular}[c]{@{}r@{}}460.13\\(-)\end{tabular}                    & \begin{tabular}[c]{@{}r@{}}487.44\\(0.94x)\end{tabular}                              & \begin{tabular}[c]{@{}r@{}}440.86\\(1.04x)\end{tabular}                              & \begin{tabular}[c]{@{}r@{}}417.31\\(1.10x)\end{tabular}           \\
                           & Model                                                  & \begin{tabular}[c]{@{}r@{}}24534.50\\(-)\end{tabular}                  & \begin{tabular}[c]{@{}r@{}}25389.74\\(0.97x)\end{tabular}                            & \begin{tabular}[c]{@{}r@{}}23712.89\\(1.03x)\end{tabular}                            & \begin{tabular}[c]{@{}r@{}}22815.79\\(1.08x)\end{tabular}         \\ 
    \midrule
    \multirow{2}{*}{LLM-L} & Attn                                                   & \begin{tabular}[c]{@{}r@{}}568.18\\(-)\end{tabular}                    & \begin{tabular}[c]{@{}r@{}}610.67\\(0.93x)\end{tabular}                              & \begin{tabular}[c]{@{}r@{}}558.56\\(1.02x)\end{tabular}                              & \begin{tabular}[c]{@{}r@{}}551.60\\(1.03x)\end{tabular}           \\
                           & Model                                                  & \begin{tabular}[c]{@{}r@{}}77378.44\\(-)\end{tabular}                  & \begin{tabular}[c]{@{}r@{}}79671.34\\(0.97x)\end{tabular}                            & \begin{tabular}[c]{@{}r@{}}75055.69\\(1.03x)\end{tabular}                            & \begin{tabular}[c]{@{}r@{}}74864.71\\(1.03x)\end{tabular}         \\
    \bottomrule
    \end{tabular}
\end{table}

This section evaluates how \name's workload-balanced context parallelism (\S\ref{sec:context_parallelism}) distributes non-causal attention execution well.
We run LLMs with 64k sequence length, where the attention mask is simulated to represent a mixed of multiple modalities.
See Appendix~\ref{sec:apdx-context-parallelism} for results with different sequence lengths.

Table~\ref{tab:context_parallel} shows the results of a single attention layer and the entire LLM with various context parallelization policies.
We only show the results of LLM-L, as the same patterns are observed in other model sizes.
\name shows the best performance, outperforming the existing causal context parallelism optimized for LLMs by up to $1.18\times$.
Intra-GPU workload balancing also shows improvement.
Even with additional overheads from aggregation, parallelizing attention subblocks within a single GPU effectively removes tail latency caused by stragglers.

\begin{figure}[t]
    \centering
    \scriptsize
    \begin{subfigure}[t]{\columnwidth}
        \includegraphics[width=\columnwidth]{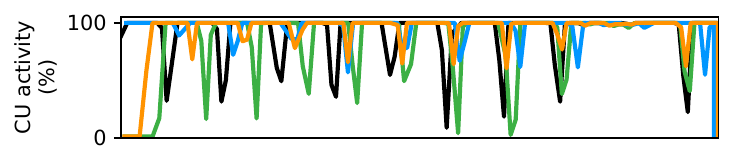}
        \caption{Causal context parallelism.}
        \label{fig:sm_analysis_zigzag}
    \end{subfigure}
    \begin{subfigure}[t]{\columnwidth}
        \includegraphics[width=\columnwidth]{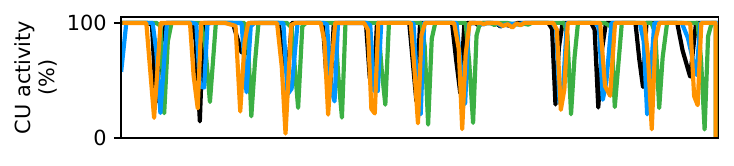}
        \caption{Inter-GPU balanced only context parallelism.}
        \label{fig:sm_analysis_inter_gpu}
    \end{subfigure}
    \begin{subfigure}[t]{\columnwidth}
        \includegraphics[width=\columnwidth]{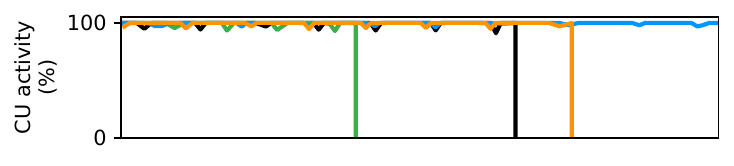}
        \caption{Intra-GPU balanced only context parallelism.}
        \label{fig:sm_analysis_intra_gpu}
    \end{subfigure}
    \begin{subfigure}[t]{\columnwidth}
        \includegraphics[width=\columnwidth]{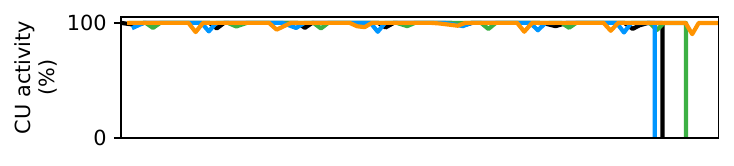}
        \caption{\name context parallelism.}
        \label{fig:sm_analysis_cornstarch}
    \end{subfigure}
    \caption{CU activity analysis with various context parallelization policies running a single attention layer of LLM-L. Each line represents one GPU.}
    \label{fig:context_parallelism_sm_analysis}
\end{figure}

Surprisingly, however, balancing workload distribution only at a token level (inter-GPU balancing only) does not provide performance improvement.
To understand this, we further perform CU activity analysis of a single attention layer, depicted in Figure~\ref{fig:context_parallelism_sm_analysis}.
Severe downward spikes are observed in both causal context parallelism (Figure~\ref{fig:sm_analysis_zigzag}) in inter-GPU only balance context parallelism (Figure~\ref{fig:sm_analysis_inter_gpu}).
The spikes happen at the end of every attention head computation.
This is because attention kernel for the next head cannot be launched until the in-flight attention kernel for the current head is entirely finished, leaving CUs inactive.
\footnotetext{A simpler alternative solution is to use multiple CUDA streams to overlap attention computations across heads. However, our experiments shows that it does not solve the CU underutilization problem. See Appendix~\ref{sec:apdx-multistream}.}
Intra-GPU balancing fundamentally solves this problem by distributing workloads of attention computation of each single block in finer granularity across CUs, showing higher CU activity (Figure~\ref{fig:sm_analysis_intra_gpu}).
Still, only balancing Intra-GPU workloads does not balance the total amount of workloads across GPUs; some GPUs become idle much earlier while others are busy, reducing overall utilization.
Combining inter- and intra-GPU balancing, \name achieves the best performance (Figure~\ref{fig:sm_analysis_cornstarch}).

\subsection{Comparison with DCP}
While DCP~\cite{dcp-sosp25} is not designed for MLLMs, its capabilities can be adapted to MLLMs, representing MLLM-specific attention patterns.
We compare \name with DCP in terms of single attention layer performance and end-to-end training time on 16 GPUs.
We use 4 attention patterns demonstrated in the DCP paper for comparison.

\begin{table}[t]
    \centering
    \small
    \caption{\name and DCP comparison using Lambda attention pattern. Time is in milliseconds.}
    \label{tab:dcp_comparison_single_attention_layer}
    \begin{tabular}{c|rrrr} 
    \toprule
    Seq length & \multicolumn{1}{c}{32k} & \multicolumn{1}{c}{64k} & \multicolumn{1}{c}{128k} & \multicolumn{1}{c}{256k}  \\ 
    \midrule
    \name & 177.41                  & 143.56                  & 375.49                   & 1193.32                   \\
    DCP        & 46.36                   & 124.72                  & 417.07                   & 1545.83                   \\
    \bottomrule
    \end{tabular}
\end{table}

\begin{table}[t]
    \centering
    \scriptsize
    \caption{End-to-end training time on \name and DCP.}
    \label{tab:dcp_comparison_end_to_end_training_time}
    \begin{tabular}{cc|rrrr} 
    \toprule
    \begin{tabular}[c]{@{}c@{}}Seq\\Length\end{tabular} & Time (s)   & \multicolumn{1}{c}{Causal} & \multicolumn{1}{c}{Lambda} & \multicolumn{1}{c}{\begin{tabular}[c]{@{}c@{}}Causal\\Blockwise\end{tabular}} & \multicolumn{1}{c}{\begin{tabular}[c]{@{}c@{}}Shared\\Questions\end{tabular}}  \\ 
    \midrule
    \multirow{2}{*}{32k}                                & \name & 5.7                        & 5.8                        & 10.5                                                                          & 5.7                                                                            \\ 
                                                        & DCP        & 26.5                       & 49.0                       & 42.0                                                                          & 46.5                                                                           \\ 
    \midrule
    \multirow{2}{*}{128k}                               & \name & 37.2                       & 37.4                       & 29.0                                                                          & 33.9                                                                           \\ 
                                                        & DCP        & 46.5                       & 107.1                      & 100.8                                                                         & 102.4                                                                          \\
    \bottomrule
    \end{tabular}
\end{table}

Table~\ref{tab:dcp_comparison_single_attention_layer} shows the result of a single attention layer with lambda attention pattern.
See Appendix~\ref{sec:apdx-dcp-comparison} for other patterns' result.
With shorter sequence length, DCP shows better performance than \name.
This is because \name requires enough sequence length to fully hide the communication overhead.
With longer sequence length, however, DCP rather struggles to hide the communication overhead, since it relies on ring context parallelism, which is inefficient in training with long sequence lengths or large number of context parallel degrees, while \name all-gather based context parallelism gets more benefits from longer sequences.

Table~\ref{tab:dcp_comparison_end_to_end_training_time} shows a full iteration time of Llama3-1b model (planning + 1 forward + 1 backward).
DCP suffers from significant planning overhead to find the optimal computation schedule.
\name's heuristic token assignment is efficient, finishing planning in less than 1 second in all cases, outperforming DCP.

\section{Related Work}
\label{sec:related_work}

\parabf{4D parallelism.}
Large-scale LLM training combines tensor, pipeline, data, and context parallelism to scale model and sequence length~\cite{megatronlm-sc21,llama3-arxiv24,megascale-nsdi24}.
Each dimension has been optimized for pipeline stage balancing and causal context parallelism~\cite{pipedream-icml21,wlbllm-osdi25,loongtrain-corr24}.
However, these methods assume homogeneous transformers and fixed causal attention, mostly focusing on LLMs, and do not address MLLM-specific heterogeneity, frozen status, or non-causal attention.

\parabf{Distributed multimodal training.}
MLLM training commonly uses FSDP~\cite{pytorchfsdp-arxiv23,llava-neurips23}, but it scales poorly for large models in large clusters~\cite{torchtitan-iclr25}.
DistMM~\cite{distmm-nsdi24}, DistTrain~\cite{disttrain-sigcomm25}, Optimus~\cite{optimus-atc25}, and DIP~\cite{dip-asplos26} address heterogeneous modality modules, yet overlook frozen-status effects on pipeline balance and non-causal context parallelism.
% \name extends 4D parallelism with frozen status-aware pipeline partitioning and workload-balanced context parallelism for MLLMs.

\section{Conclusion}
\label{sec:conclusion}

In this paper, we presented \name, a multimodality-aware distributed MLLM training framework.
% We observe that simply retrofitting existing distributed training frameworks for MLLMs is not sufficient.
% Instead, \name addresses challenges arising from model and data heterogeneity in MLLM training from first principles.
\name addresses higher-order challenges arising from model and data heterogeneity in MLLM training.
% We introduce a new parallelism dimension called modality parallelism that allows for flexible scheduling of modality encoders.
We introduce frozen status-aware pipeline parallelism that balances the computational cost of MLLM pipeline stages.
We also introduce workload balanced context parallelism which computes the amount of workloads both in intra-GPU and inter-GPU.
\name provides $2.26\times$ speedup over the state-of-the-art distributed MLLM training frameworks on average.

\label{EndOfPaper}

\newpage

\section*{Acknowledgements}
We thank the ICML reviewers and members of SymbioticLab for their helpful discussions and feedback.
This work was supported in part by NSF grants CCF-2450085, CCF-2327011, CCF-2504995, CNS-2106184, CNS-2535540, and CNS-2406598 and by grants from Cisco, Ford, Mozilla Foundation, and Laude Institute.

\section*{Impact Statement}

This work improves the efficiency of distributed MLLM training.
Considering the growing demand for multimodal LLMs, our work is expected to have a significant time and energy savings in training large multimodal models.

% Authors are \textbf{required} to include a statement of the potential broader
% impact of their work, including its ethical aspects and future societal
% consequences. This statement should be in an unnumbered section at the end of
% the paper (co-located with Acknowledgements -- the two may appear in either
% order, but both must be before References), and does not count toward the paper
% page limit. In many cases, where the ethical impacts and expected societal
% implications are those that are well established when advancing the field of
% Machine Learning, substantial discussion is not required, and a simple
% statement such as the following will suffice:

% ``This paper presents work whose goal is to advance the field of Machine
% Learning. There are many potential societal consequences of our work, none
% which we feel must be specifically highlighted here.''

% The above statement can be used verbatim in such cases, but we encourage
% authors to think about whether there is content which does warrant further
% discussion, as this statement will be apparent if the paper is later flagged
% for ethics review.

% % In the unusual situation where you want a paper to appear in the
% % references without citing it in the main text, use \nocite

\bibliography{ref}

@inproceedings{megatronlm-sc21,
    title = {Efficient large-scale language model training on GPU clusters using megatron-LM},
    author = {Narayanan, Deepak and Shoeybi, Mohammad and Casper, Jared and LeGresley, Patrick and Patwary, Mostofa and Korthikanti, Vijay and Vainbrand, Dmitri and Kashinkunti, Prethvi and Bernauer, Julie and Catanzaro, Bryan and Phanishayee, Amar and Zaharia, Matei},
    booktitle = {SC},
    year = {2021},
}

@misc{megatronlm-github,
    title = {Megatron-LM},
    author = {NVIDIA},
    year = {2024},
    url = {https://github.com/NVIDIA/Megatron-LM}
}

@inproceedings{pytorch-neurips19,
    title = {PyTorch: An Imperative Style, High-Performance Deep Learning Library},
    author = {Paszke, Adam and Gross, Sam and Massa, Francisco and Lerer, Adam and Bradbury, James and Chanan, Gregory and Killeen, Trevor and Lin, Zeming and Gimelshein, Natalia and Antiga, Luca and Desmaison, Alban and Kopf, Andreas and Yang, Edward and DeVito, Zachary and Raison, Martin and Tejani, Alykhan and Chilamkurthy, Sasank and Steiner, Benoit and Fang, Lu and Bai, Junjie and Chintala, Soumith},
    booktitle = {NeurIPS},
    year = {2019},
}

@inproceedings{colossalai-icpp23,
    title = {Colossal-AI: A Unified Deep Learning System For Large-Scale Parallel Training},
    author = {Li, Shenggui and Liu, Hongxin and Bian, Zhengda and Fang, Jiarui and Huang, Haichen and Liu, Yuliang and Wang, Boxiang and You, Yang},
    booktitle = {ICPP},
    year = {2023},
}

@inproceedings{torchtitan-iclr25,
title={TorchTitan: One-stop PyTorch native solution for production ready {LLM} pretraining},
author={Wanchao Liang and Tianyu Liu and Less Wright and Will Constable and Andrew Gu and Chien-Chin Huang and Iris Zhang and Wei Feng and Howard Huang and Junjie Wang and Sanket Purandare and Gokul Nadathur and Stratos Idreos},
booktitle={ICLR},
year={2025},
url={https://openreview.net/forum?id=SFN6Wm7YBI}
}

@inproceedings{deepspeed-sc20,
    title = {DeepSpeed: System Optimizations Enable Training Deep Learning Models with Over 100 Billion Parameters},
    author = {Rasley, Jeff and Rajbhandari, Samyam and Ruwase, Olatunji and He, Yuxiong},
    booktitle = {KDD},
    year = {2020},
}

@inproceedings{transformers-acl20,
    title = {Transformers: State-of-the-Art Natural Language Processing},
    author = {Wolf, Thomas and Debut, Lysandre and Sanh, Victor and Chaumond, Julien and Delangue, Clement and Moi, Anthony and Cistac, Pierric and Rault, Tim and Louf, Remi and Funtowicz, Morgan and Davison, Joe and Shleifer, Sam and von Platen, Patrick and Ma, Clara and Jernite, Yacine and Plu, Julien and Xu, Canwen and Le Scao, Teven and Gugger, Sylvain and Drame, Mariama and Lhoest, Quentin and Rush, Alexander},
    booktitle = {EMNLP},
    year = {2020},
}

@misc{llama3-arxiv24,
    title = {The Llama 3 Herd of Models},
    author = {Meta AI},
    year = {2024},
    url = {https://arxiv.org/abs/2407.21783}
}

@inproceedings{oobleck-sosp23,
    title = {Oobleck: Resilient Distributed Training of Large Models Using Pipeline Templates},
    author = {Jang, Insu and Yang, Zhenning and Zhang, Zhen and Jin, Xin and Chowdhury, Mosharaf},
    booktitle = {SOSP},
    year = {2023},
}

@inproceedings{graphpipe-asplos25,
    author = {Jeon, Byungsoo and Wu, Mengdi and Cao, Shiyi and Kim, Sunghyun and Park, Sunghyun and Aggarwal, Neeraj and Unger, Colin and Arfeen, Daiyaan and Liao, Peiyuan and Miao, Xupeng and Alizadeh, Mohammad and Ganger, Gregory R. and Chen, Tianqi and Jia, Zhihao},
    title = {GraphPipe: Improving Performance and Scalability of DNN Training with Graph Pipeline Parallelism},
    year = {2025},
    url = {https://doi.org/10.1145/3669940.3707220},
    doi = {10.1145/3669940.3707220},
    booktitle = {ASPLOS}
}

@inproceedings{dip-asplos26,
author = {Xue, Zhenliang and Hu, Hanpeng and Chen, Xing and Jiang, Yimin and Song, Yixin and Mi, Zeyu and Zhu, Yibo and Jiang, Daxin and Xia, Yubin and Chen, Haibo},
title = {DIP: Efficient Large Multimodal Model Training with Dynamic Interleaved Pipeline},
year = {2026},
doi = {10.1145/3779212.3790154},
url = {https://doi.org/10.1145/3779212.3790154},
booktitle = {ASPLOS},
}

@inproceedings{llama3training-isca25,
author = {Chu, Weiwei and Xie, Xinfeng and Yu, Jiecao and Wang, Jie and Phanishayee, Amar and Tang, Chunqiang and Hao, Yuchen and Huang, Jianyu and Ozdal, Mustafa and Wang, Jun and Goswami, Vedanuj and Goyal, Naman and Kadian, Abhishek and Gu, Andrew and Cai, Chris and Tian, Feng and Wang, Xiaodong and Si, Min and Balaji, Pavan and Chu, Ching-Hsiang and Park, Jongsoo},
title = {Scaling Llama 3 Training with Efficient Parallelism Strategies},
year = {2025},
booktitle = {ISCA},
}

@inproceedings{dcp-sosp25,
author = {Jiang, Chenyu and Cai, Zhenkun and Tian, Ye and Jia, Zhen and Wang, Yida and Wu, Chuan},
title = {DCP: Addressing Input Dynamism In Long-Context Training via Dynamic Context Parallelism},
year = {2025},
booktitle = {SOSP},
}

@inproceedings{alpa-osdi22,
    title = {Alpa: Automating Inter- and Intra-Operator Parallelism for Distributed Deep Learning},
    author = {Zheng, Lianmin and Li, Zhuohan and Zhang, Hao and Zhuang, Yonghao and Chen, Zhifeng and Huang, Yanping and Wang, Yida and Xu, Yuanzhong and Zhuo, Danyang and Xing, Eric P. and Gonzalez, Joseph E. and Stoica, Ion},
    booktitle = {OSDI},
    year = {2022},
}

@inproceedings{adapipe-asplos24,
    author = {Sun, Zhenbo and Cao, Huanqi and Wang, Yuanwei and Feng, Guanyu and Chen, Shengqi and Wang, Haojie and Chen, Wenguang},
    title = {AdaPipe: Optimizing Pipeline Parallelism with Adaptive Recomputation and Partitioning},
    year = {2024},
    url = {https://doi.org/10.1145/3620666.3651359},
    doi = {10.1145/3620666.3651359},
    series = {ASPLOS}
}

@article{galvatron-vldb22,
    author = {Miao, Xupeng and Wang, Yujie and Jiang, Youhe and Shi, Chunan and Nie, Xiaonan and Zhang, Hailin and Cui, Bin},
    title = {Galvatron: Efficient Transformer Training over Multiple GPUs Using Automatic Parallelism},
    year = {2022},
    issue_date = {November 2022},
    publisher = {VLDB Endowment},
    volume = {16},
    number = {3},
    issn = {2150-8097},
    url = {https://doi.org/10.14778/3570690.3570697},
    doi = {10.14778/3570690.3570697},
    journal = {VLDB},
    month = nov,
    pages = {470-479},
    numpages = {10}
}

@misc{pytorchfsdp-arxiv23,
      title={PyTorch FSDP: Experiences on Scaling Fully Sharded Data Parallel}, 
      author={Yanli Zhao and Andrew Gu and Rohan Varma and Liang Luo and Chien-Chin Huang and Min Xu and Less Wright and Hamid Shojanazeri and Myle Ott and Sam Shleifer and Alban Desmaison and Can Balioglu and Pritam Damania and Bernard Nguyen and Geeta Chauhan and Yuchen Hao and Ajit Mathews and Shen Li},
      year={2023},
      eprint={2304.11277},
      archivePrefix={arXiv},
      primaryClass={cs.DC},
      url={https://arxiv.org/abs/2304.11277}, 
}

@inproceedings{flashattn-nips22,
    title = {FlashAttention: Fast and Memory-Efficient Exact Attention with IO-Awareness},
    author = {Dao, Tri and Fu, Daniel Y. and Ermon, Stefano and Rudra, Atri and R{\'e}, Christopher},
    booktitle = {NeurIPS},
    year = {2022},
}

@inproceedings{flashattn2-iclr24,
    title = {Flash{A}ttention-2: Faster Attention with Better Parallelism and Work Partitioning},
    author = {Dao, Tri},
    booktitle = {ICLR},
    year = {2024}
}

@misc{onlinesoftmax-arxiv18,
    title={Online normalizer calculation for softmax}, 
    author={Maxim Milakov and Natalia Gimelshein},
    year={2018},
    eprint={1805.02867},
    archivePrefix={arXiv},
    primaryClass={cs.PF},
    url={https://arxiv.org/abs/1805.02867}, 
}

@misc{memory-efficient-attention-arxiv21,
    title = {Self-attention Does Not Need $O(n^2)$ Memory},
    author = {Rabe, Markus N. and Staats, Charles},
    year = {2022},
    url = {https://arxiv.org/abs/2112.05682}
}

@inproceedings{flexattention-mlsys25,
  title={FlexAttention: A Programming Model for Generating Fused Attention Variants},
  author={Dong, Juechu and Feng, Boyuan and Guessous, Driss and He, Horace},
  booktitle={MLSys},
  year={2025}
}

@inproceedings{flashmask-iclr25,
title={FlashMask: Efficient and Rich Mask Extension of FlashAttention},
author={Guoxia Wang and Jinle Zeng and Xiyuan Xiao and Siming Wu and Jiabin Yang and Lujing Zheng and Zeyu Chen and Jiang Bian and Dianhai Yu and Haifeng Wang},
booktitle={The Thirteenth International Conference on Learning Representations},
year={2025},
url={https://openreview.net/forum?id=wUtXB43Chi}
}

@inproceedings{triton-mapl19,
    author = {Tillet, Philippe and Kung, Hsiang-Tsung and Cox, David},
    title = {Triton: an intermediate language and compiler for tiled neural network computations},
    year = {2019},
    isbn = {9781450367196},
    publisher = {Association for Computing Machinery},
    address = {New York, NY, USA},
    url = {https://doi.org/10.1145/3315508.3329973},
    booktitle = {MAPL},
}

@misc{triton-github21,
    title = {Triton: Open-source GPU Programming for Neural Networks},
    author = {OpenAI},
    year = {2021},
    url = {https://github.com/triton-lang/triton}
}

@inproceedings{contextparallel-mlsys25,
title={Context Parallelism for Scalable Million-Token Inference},
author={Amy Yang and Jingyi Yang and Aya Ibrahim and Xinfeng Xie and Bangsheng Tang and Grigory Sizov and Jongsoo Park and Jianyu Huang},
booktitle={Eighth Conference on Machine Learning and Systems},
year={2025},
url={https://openreview.net/forum?id=Vmf09yVJhT}
}

@inproceedings{ringattention-iclr24,
    title = {RingAttention with Blockwise Transformers for Near-Infinite Context},
    author = {Liu, Hao and Zaharia, Matei and Abbeel, Pieter},
    booktitle = {ICLR},
    year = {2024}
}

@misc{ulysses-arxiv23,
    title = {DeepSpeed Ulysses: System Optimizations for Enabling Training of Extreme Long Sequence Transformer Models},
    author = {Jacobs, Sam Ade and Tanaka, Masahiro and Zhang, Chengming and Zhang, Minjia and Song, Shuaiwen Leon and Rajbhandari, Samyam and He, Yuxiong},
    year = {2023},
    url = {https://arxiv.org/abs/2309.14509}
}

@article{loongtrain-corr24,
    title = {LoongTrain: Efficient Training of Long-Sequence LLMs with Head-Context Parallelism},
    author = {Gu, Diandian and Sun, Peng and Hu, Qinghao and Huang, Ting and Chen, Xun and Xiong, Yingtong and Wang, Guoteng and Chen, Qiaoling and Zhao, Shangchun and Fang, Jiarui and Wen, Yonggang and Zhang, Tianwei and Jin, Xin and Liu, Xuanzhe},
    journal = {CoRR},
    volume = {abs/2406.18485},
    year = {2024},
    url = {https://doi.org/10.48550/arXiv.2406.18485}
}

@InProceedings{pipedream-icml21,
  title = 	 {Memory-Efficient Pipeline-Parallel DNN Training},
  author =       {Narayanan, Deepak and Phanishayee, Amar and Shi, Kaiyu and Chen, Xie and Zaharia, Matei},
  booktitle = 	 {ICML},
  year = 	 {2021},
  url = 	 {https://proceedings.mlr.press/v139/narayanan21a.html},
}

@inproceedings{zerobubblepp-iclr24,
    title = {Zero Bubble (Almost) Pipeline Parallelism},
    author = {Qi, Penghui and Wan, Xinyi and Huang, Guangxing and Lin, Min},
    booktitle = {ICLR},
    year = {2024}
}

@inproceedings {distmm-nsdi24,
    author = {Jun Huang and Zhen Zhang and Shuai Zheng and Feng Qin and Yida Wang},
    title = {{DistMM}: Accelerating Distributed Multimodal Model Training},
    booktitle = {NSDI},
    year = {2024},
    url = {https://www.usenix.org/conference/nsdi24/presentation/huang},
}

@inproceedings{optimus-atc25,
    title={Optimus: Accelerating Large-Scale Multi-Modal LLM Training by Bubble Exploitation},
    author={Feng, Weiqi and Chen, Yangrui and Wang, Shaoyu and Peng, Yanghua and Lin, Haibin and Yu, Minlan},
    booktitle={USENIX ATC},
    pages={161--177},
    year={2025}
}

@inproceedings{mm-llms-acl24,
    title = {MM-LLMs: Recent Advances in MultiModal Large Language Models},
    author = {Zhang, Duzhen and Yu, Yahan and Dong, Jiahua and Li, Chenxing and Su, Dan and Chu, Chenhui and Yu, Dong},
    booktitle = {ACL Findings},
    year = {2024},
}

@inproceedings{languagebind-iclr24,
    title = {LanguageBind: Extending Video-Language Pretraining to N-modality by Language-based Semantic Alignment},
    author = {Zhu, Bin and Lin, Bin and Ning, Munan and Yan, Yang and Cui, Jiaxi and Wang, HongFa and Pang, Yatian and Jiang, Wenhao and Zhang, Junwu and Li, Zongwei and Zhang, Cai Wan and Li, Zhifeng and Liu, Wei and Yuan, Li},
    booktitle = {ICLR},
    year = {2024}
}

@inproceedings{llava-neurips23,
    title = {Visual Instruction Tuning},
    author = {Liu, Haotian and Li, Chunyuan and Wu, Qingyang and Lee, Yong Jae},
    booktitle = {NeurIPS},
    year = {2023},
}

@inproceedings{llavanext-cvpr24,
    title = {Improved Baselines with Visual Instruction Tuning},
    author = {Liu, Haotian and Li, Chunyuan and Li, Yuheng and Lee, Yong Jae},
    booktitle = {CVPR},
    year = {2024},
}

@inproceedings{internvl-cvpr24,
    title = {InternVL: Scaling up Vision Foundation Models and Aligning for Generic Visual-Linguistic Tasks},
    author = {Chen, Zhe and Wu, Jiannan and Wang, Wenhai and Su, Weijie and Chen, Guo and Xing, Sen and Zhong, Muyan and Zhang, Qinglong and Zhu, Xizhou and Lu, Lewei and Li, Bin and Luo, Ping and Lu, Tong and Qiao, Yu and Dai, Jifeng},
    booktitle = {CVPR},
    year = {2024},
}

@misc{qwen2vl-arxiv24,
    title = {Qwen2-VL: Enhancing Vision-Language Model's Perception of the World at Any Resolution},
    author = {Wang, Peng and Bai, Shuai and Tan, Sinan and Wang, Shijie and Fan, Zhihao and Bai, Jinze and others},
    year = {2024},
    url = {https://arxiv.org/abs/2409.12191}
}

@InProceedings{vila-cvpr24,
    author    = {Lin, Ji and Yin, Hongxu and Ping, Wei and Molchanov, Pavlo and Shoeybi, Mohammad and Han, Song},
    title     = {VILA: On Pre-training for Visual Language Models},
    booktitle = {CVPR},
    month     = {June},
    year      = {2024},
}

@inproceedings{vlm-acl21,
    title = {VLM: Task-agnostic Video-Language Model Pre-training for Video Understanding},
    author = {Xu, Hu and Ghosh, Gargi and Huang, Po-Yao and Arora, Prahal and Aminzadeh, Masoumeh and Feichtenhofer, Christoph and Metze, Florian and Zettlemoyer, Luke},
    booktitle = {ACL Findings},
    year = {2021},
}

@misc{phi4-arxiv24,
      title={Phi-4 Technical Report}, 
      author={Marah Abdin and Jyoti Aneja and Harkirat Behl and Sébastien Bubeck and Ronen Eldan and Suriya Gunasekar and Michael Harrison and Russell J. Hewett and Mojan Javaheripi and Piero Kauffmann and James R. Lee and Yin Tat Lee and Yuanzhi Li and Weishung Liu and Caio C. T. Mendes and Anh Nguyen and Eric Price and Gustavo de Rosa and Olli Saarikivi and Adil Salim and Shital Shah and Xin Wang and Rachel Ward and Yue Wu and Dingli Yu and Cyril Zhang and Yi Zhang},
      year={2024},
      eprint={2412.08905},
      archivePrefix={arXiv},
      primaryClass={cs.CL},
      url={https://arxiv.org/abs/2412.08905}, 
}

@misc{videollava-arxiv24,
    title = {Video-LLaVA: Learning United Visual Representation by Alignment Before Projection},
    author = {Lin, Bin and Ye, Yang and Zhu, Bin and Cui, Jiaxi and Ning, Munan and Jin, Peng and Yuan, Li},
    year = {2024},
    url = {https://arxiv.org/abs/2311.10122}
}

@misc{cogvlm-arxiv24,
      title={CogVLM2: Visual Language Models for Image and Video Understanding}, 
      author={Wenyi Hong and Weihan Wang and Ming Ding and Wenmeng Yu and Qingsong Lv and Yan Wang and Yean Cheng and Shiyu Huang and Junhui Ji and Zhao Xue and Lei Zhao and Zhuoyi Yang and Xiaotao Gu and Xiaohan Zhang and Guanyu Feng and Da Yin and Zihan Wang and Ji Qi and Xixuan Song and Peng Zhang and Debing Liu and Bin Xu and Juanzi Li and Yuxiao Dong and Jie Tang},
      year={2024},
      eprint={2408.16500},
      archivePrefix={arXiv},
      primaryClass={cs.CV},
      url={https://arxiv.org/abs/2408.16500}, 
}

@inproceedings{ll3da-cvpr24,
    title = {LL3DA: Visual Interactive Instruction Tuning for Omni-3D Understanding Reasoning and Planning},
    author = {Chen, Sijin and Chen, Xin and Zhang, Chi and Li, Mingsheng and Yu, Gang and Fei, Hao and Zhu, Hongyuan and Fan, Jiayuan and Chen, Tao},
    booktitle = {CVPR},
    year = {2024},
}

@inproceedings{vlm-understanding-neurips24,
    title={Building and better understanding vision-language models: insights and future directions},
    author={Hugo Lauren{\c{c}}on and Andr{\'e}s Marafioti and Victor Sanh and Leo Tronchon},
    booktitle={Workshop on Responsibly Building the Next Generation of Multimodal Foundational Models},
    year={2024},
    url={https://openreview.net/forum?id=iSL0FHZStr}
}

@InProceedings{survey-vlm-cvpr25,
    author    = {Li, Zongxia and Wu, Xiyang and Du, Hongyang and Liu, Fuxiao and Nghiem, Huy and Shi, Guangyao},
    title     = {A Survey of State of the Art Large Vision Language Models: Benchmark Evaluations and Challenges},
    booktitle = {Proceedings of the IEEE/CVF Conference on Computer Vision and Pattern Recognition (CVPR) Workshops},
    month     = {June},
    year      = {2025},
    pages     = {1587-1606}
}

@article{cambrian-corr24,
    title = {Cambrian-1: A Fully Open, Vision-Centric Exploration of Multimodal LLMs},
    author = {Tong, Shengbang and Brown, Ellis and Wu, Penghao and Woo, Sanghyun and Middepogu, Manoj and Akula, Sai Charitha and Yang, Jihan and Yang, Shusheng and Iyer, Adithya and Pan, Xichen and Wang, Austin and Fergus, Rob and LeCun, Yann and Xie, Saining},
    journal = {CoRR},
    volume = {abs/2406.16860},
    year = {2024},
    url = {https://doi.org/10.48550/arXiv.2406.16860}
}

@misc{gemma-arxiv24,
    title = {Gemma: Open Models Based on Gemini Research and Technology},
    author = {Google Deepmind},
    year = {2024},
    url = {https://arxiv.org/abs/2403.08295}
}

@misc{gemma2-arxiv24,
    title = {Gemma 2: Improving Open Language Models at a Practical Size},
    author = {Google Deepmind},
    year = {2024},
    url = {https://arxiv.org/abs/2408.00118}
}

@misc{gemma3-arxiv25,
      title={Gemma 3 Technical Report}, 
      author={Google Deepmind},
      year={2025},
      eprint={2503.19786},
      archivePrefix={arXiv},
      primaryClass={cs.CL},
      url={https://arxiv.org/abs/2503.19786}, 
}

@article{internlm2-corr24,
    title = {InternLM2 Technical Report},
    author = {Zheng Cai and Maosong Cao and Haojiong Chen and Kai Chen and Keyu Chen and Xin Chen and Xun Chen and Zehui Chen and Zhi Chen and Pei Chu and others},
    journal = {CoRR},
    volume = {abs/2403.17297},
    year = {2024},
    url = {https://doi.org/10.48550/arXiv.2403.17297}
}

@misc{mistral-arxiv23,
    title = {Mistral 7B},
    author = {Jiang, Albert Q. and Sablayrolles, Alexandre and Mensch, Arthur and Bamford, Chris and Chaplot, Devendra Singh and de las Casas, Diego and others},
    year = {2023},
    url = {https://arxiv.org/abs/2310.06825}
}

@misc{mixtral-arxiv24,
    title = {Mixtral of Experts},
    author = {Jiang, Albert Q. and Sablayrolles, Alexandre and Roux, Antoine and Mensch, Arthur and Savary, Blanche and Bamford, Chris and others},
    year = {2024},
    url = {https://arxiv.org/abs/2401.04088}
}

@misc{pixtral-arxiv24,
    title = {Pixtral: A Large-Scale Vision-Language Model with Parameter-Efficient Fine-Tuning},
    author = {Agrawal, Pravesh and Antoniak, Szymon and Bou Hanna, Emma and Bout, Baptiste and Chaplot, Devendra and others},
    year = {2024},
    url = {https://arxiv.org/abs/2410.07073}
}

@misc{phi3-arxiv24,
    title = {Phi-3 Technical Report: A Highly Capable Language Model Locally on Your Phone},
    author = {Microsoft},
    year = {2024},
    url = {https://arxiv.org/abs/2404.14219}
}

@article{qwen2-corr24,
    title = {Qwen2 Technical Report},
    author = {Yang, An and Yang, Baosong and Hui, Binyuan and Zheng, Bo and Yu, Bowen and Zhou, Chang and others},
    journal = {CoRR},
    volume = {abs/2407.10671},
    year = {2024},
    url = {https://doi.org/10.48550/arXiv.2407.10671}
}

@inproceedings{clip-icml21,
    title = {Learning Transferable Visual Models From Natural Language Supervision},
    author = {Radford, Alec and Kim, Jong Wook and Hallacy, Chris and Ramesh, Aditya and Goh, Gabriel and others},
    booktitle = {ICML},
    year = {2021},
}

@article{dinov2-tmlr24,
    title = {DINOv2: Learning Robust Visual Features without Supervision},
    author = {Oquab, Maxime and Darcet, Timothée and Moutakanni, Théo and Vo, Huy V. and Szafraniec, Marc and others},
    journal = {TMLR},
    issn={2835-8856},
    year = {2024},
    url = {https://openreview.net/forum?id=a68SUt6zFt}
}

@misc{evaclip-arxiv23,
    title = {EVA-CLIP: Improved Training Techniques for CLIP at Scale},
    author = {Sun, Quan and Fang, Yuxin and Wu, Ledell and Wang, Xinlong and Cao, Yue},
    year = {2023},
    url = {https://arxiv.org/abs/2303.15389}
}

@inproceedings{siglip-iccv23,
    title = {Sigmoid Loss for Language Image Pre-Training},
    author = {Zhai, Xiaohua and Mustafa, Basil and Kolesnikov, Alexander and Beyer, Lucas},
    booktitle = {ICCV},
    year = {2023},
}

@InProceedings{blip2-icml23,
  title = 	 {{BLIP}-2: Bootstrapping Language-Image Pre-training with Frozen Image Encoders and Large Language Models},
  author =       {Li, Junnan and Li, Dongxu and Savarese, Silvio and Hoi, Steven},
  booktitle = 	 {ICML},
  pages = 	 {19730--19742},
  year = 	 {2023},
  editor = 	 {Krause, Andreas and Brunskill, Emma and Cho, Kyunghyun and Engelhardt, Barbara and Sabato, Sivan and Scarlett, Jonathan},
  volume = 	 {202},
  series = 	 {Proceedings of Machine Learning Research},
  month = 	 {23--29 Jul},
  publisher =    {PMLR},}

@misc{qwen2audio-arxiv24,
    title = {Qwen2-Audio Technical Report},
    author = {Chu, Yunfei and Xu, Jin and Yang, Qian and Wei, Haojie and Wei, Xipin and Guo, Zhifang and others},
    year = {2024},
    url = {https://arxiv.org/abs/2407.10759}
}

@inproceedings{whisper-icml23,
    title = {Robust Speech Recognition via Large-Scale Weak Supervision},
    author = {Radford, Alec and Kim, Jong Wook and Xu, Tao and Brockman, Greg and Mcleavey, Christine and Sutskever, Ilya},
    booktitle = {ICML},
    year = {2023},
}

@article{graham-scheduling-siam69,
    title = {Bounds on Multiprocessing Timing Anomalies},
    author = {Graham, Ronald Lewis},
    journal = {SIAM Journal on Applied Mathematics},
    volume = {17},
    number = {2},
    pages = {416-429},
    year = {1969},
    url = {https://doi.org/10.1137/0117039}
}

@inproceedings{actrecomp-mlsys23,
    title = {Reducing Activation Recomputation in Large Transformer Models},
    author = {Korthikanti, Vijay Anand and Casper, Jared and Lym, Sangkug and McAfee, Lawrence and Andersch, Michael and Shoeybi, Mohammad and Catanzaro, Bryan},
    booktitle = {MLSys},
    year = {2023}
}

@inproceedings{blockwise-neurip23,
    title = {Blockwise Parallel Transformers for Large Context Models},
    author = {Liu, Hao and Abbeel, Pieter},
    booktitle = {NeurIPS},
    year = {2023}
}

@inproceedings{unity-osdi22,
    title = {Unity: Accelerating DNN Training Through Joint Optimization of Algebraic Transformations and Parallelization},
    author = {Unger, Colin and Jia, Zhihao and Wu, Wei and Lin, Sina and Baines, Mandeep and others},
    booktitle = {OSDI},
    year = {2022},
}

@inproceedings{megascale-nsdi24,
    title = {MegaScale: Scaling Large Language Model Training to More Than 10,000 GPUs},
    author = {Jiang, Ziheng and Lin, Haibin and Zhong, Yinmin and Huang, Qi and Chen, Yangrui and others},
    booktitle = {NSDI},
    year = {2024},
}

@inproceedings{disttrain-sigcomm25,
      title={DistTrain: Addressing Model and Data Heterogeneity with Disaggregated Training for Multimodal Large Language Models}, 
      author={Zili Zhang and Yinmin Zhong and Ranchen Ming and Hanpeng Hu and Jianjian Sun and Zheng Ge and Yibo Zhu and Xin Jin},
      booktitle = {SIGCOMM},
      year={2025},
}

@misc{bytescale-arxiv25,
      title={ByteScale: Efficient Scaling of LLM Training with a 2048K Context Length on More Than 12,000 GPUs}, 
      author={Hao Ge and Junda Feng and Qi Huang and Fangcheng Fu and Xiaonan Nie and Lei Zuo and Haibin Lin and Bin Cui and Xin Liu},
      year={2025},
      eprint={2502.21231},
      archivePrefix={arXiv},
      primaryClass={cs.DC},
      url={https://arxiv.org/abs/2502.21231}, 
}

@inproceedings{wlbllm-osdi25,
    title = {WLB-LLM: Workload-Balanced 4D Parallelism for Large Language Model Training},
    author = {Zheng Wang and Anna Cai and Xinfeng Xie and Zaifeng Pan and Yue Guan and Weiwei Chu and Jie Wang and Shikai Li and Jianyu Huang and Chris Cai and Yuchen Hao and Yufei Ding},
    booktitle = {USENIX OSDI},
    year = {2025},
}

@article{deepfusion-survey24,
author = {Zhao, Fei and Zhang, Chengcui and Geng, Baocheng},
title = {Deep Multimodal Data Fusion},
year = {2024},
issue_date = {September 2024},
publisher = {Association for Computing Machinery},
address = {New York, NY, USA},
volume = {56},
number = {9},
issn = {0360-0300},
url = {https://doi.org/10.1145/3649447},
doi = {10.1145/3649447},
journal = {Computing Surveys},
month = apr,
articleno = {216},
numpages = {36}
}
\bibliographystyle{icml2026}

%%%%%%%%%%%%%%%%%%%%%%%%%%%%%%%%%%%%%%%%%%%%%%%%%%%%%%%%%%%%%%%%%%%%%%%%%%%%%%%
%%%%%%%%%%%%%%%%%%%%%%%%%%%%%%%%%%%%%%%%%%%%%%%%%%%%%%%%%%%%%%%%%%%%%%%%%%%%%%%
% APPENDIX
%%%%%%%%%%%%%%%%%%%%%%%%%%%%%%%%%%%%%%%%%%%%%%%%%%%%%%%%%%%%%%%%%%%%%%%%%%%%%%%
%%%%%%%%%%%%%%%%%%%%%%%%%%%%%%%%%%%%%%%%%%%%%%%%%%%%%%%%%%%%%%%%%%%%%%%%%%%%%%%
\newpage
\appendix
% \onecolumn

\appendix

\section{\name Programming Interface}
\label{sec:apdx-programming-interface}

\begin{lstlisting}[float, morekeywords={MultimodalModule,ParallelSpec,MultimodalParallelModule,execute}, caption={\name APIs for distributed MLLM training.}, label=code:api]
# Load unimodal models
vis = SiglipVisionModel.from_pretrained("...")
aud = WhisperEncoder.from_pretrained("...")
llm = LlamaForCausalLM.from_pretrained("...")

# Create an MLLM with modularity information
mllm = MultimodalModule(|\label{line:modality_module}|
  encoders = {
    "vision": EncoderModule(vis, proj="mlp"),
    "audio": EncoderModule(aud, proj="linear"),
    # ... more encoders
  },
  language_model = llm,
)

# Define parallel spec per modality
# either by manually or by automatically
vis_spec = ParallelSpec(...)|\label{line:parallel_spec_start}|
aud_spec = ParallelSpec(...)
llm_spec = ParallelSpec(...)|\label{line:parallel_spec_end}|

# Parallelize the MLLM
torch.distributed.init_process_group(...)
dist_mllm = MultimodalParallelModule(|\label{line:parallel_module}|
  mllm,
  modality_parallelism="parallel",|\label{line:modality_parallelism}|
  encoder_specs={|\label{line:add_spec_start}|
    "vision": vis_spec,
    "audio": aud_spec,
    # ... more encoders
  },
  language_model_spec=llm_spec,|\label{line:add_spec_end}|
  num_microbatches=...,
  microbatch_size=...,
)

# Run distributed training of MLLM
for batch in dataloader:
  output = dist_mllm.execute(batch)|\label{line:execute}|
  optimizer.step()
  optimizer.zero_grad()
\end{lstlisting}

Listing~\ref{code:api} shows the programming interface of \name.
MultimodalModule is a wrapper class that contains the modality encoders and the LLM that can be executed standalone without parallelization specifications (line~\ref{line:modality_module}).
\name accepts parallelization specifications for each modality encoder and the LLM (line~\ref{line:parallel_spec_start} to line~\ref{line:parallel_spec_end}).
The parallelization specifications are passed to MultimodalParallelModule, which is a wrapper class that contains the specification and more hyperparameters required for distributed training (line~\ref{line:parallel_module}).
After creating a distributed MLLM, users can call \texttt{execute} method to run the training (line~\ref{line:execute}).

\section{Supported Models}
\label{sec:apdx-model-list}

\begin{table}[t]
  \centering
  \scriptsize
  \caption{Supported models in \name.}
  \label{tab:apdx_supported_models}
  \begin{tabularx}{\columnwidth}{cX}
    \toprule
    Modality                                           & \multicolumn{1}{c}{Model Names}                                                        \\ 
    \midrule
    LLM                                                     & Llama (3, 4)~\cite{llama3-arxiv24}, Mistral~\cite{mistral-arxiv23}, Mixtral~\cite{mixtral-arxiv24}, Gemma (1, 2)~\cite{gemma-arxiv24, gemma2-arxiv24, gemma3-arxiv25}, Qwen (2, 2.5, 3)~\cite{qwen2-corr24}, Phi (3, 4)~\cite{phi3-arxiv24}, InternLM2~\cite{internlm2-corr24}  \\
    \midrule
    \makecell{Vision\\Encoder} & CLIP~\cite{clip-icml21}, Dinov2~\cite{dinov2-tmlr24}, Siglip~\cite{siglip-iccv23}, EvaCLIP~\cite{evaclip-arxiv23}, Pixtral~\cite{pixtral-arxiv24}, Qwen2Vision~\cite{qwen2vl-arxiv24}                                    \\
    \midrule
    \makecell{Audio\\Encoder}  & Whisper~\cite{whisper-icml23}, Qwen2Audio~\cite{qwen2audio-arxiv24}, Phi4Audio~\cite{phi4-arxiv24}                                                         \\
    \bottomrule
  \end{tabularx}
\end{table}

Table~\ref{tab:apdx_supported_models} lists the supported models in \name at the time of writing.
However, \name is not limited to these models; those in the table are just the ones we have tested.

\section{Attention Patterns}
\label{sec:apdx-attention-patterns}

\begin{figure}[t]
  \centering
  \begin{subfigure}[t]{0.3\columnwidth}
    \includegraphics[width=\columnwidth]{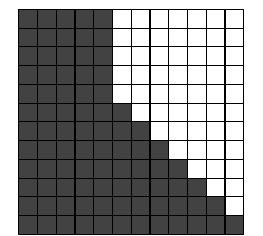}
    \caption{Encoder outputs prepended. Early VLMs used~\cite{llava-neurips23}.}
    \label{fig:attention_mask1}
  \end{subfigure}
  \hfill
  \begin{subfigure}[t]{0.3\columnwidth}
    \includegraphics[width=\columnwidth]{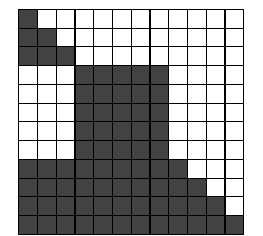}
    \caption{Encoder outputs embedded. Modern VLMs use a special token such as <image> to specify the location of modality tokens.}
    \label{fig:attention_mask2}
  \end{subfigure}
  \hfill
  \begin{subfigure}[t]{0.3\columnwidth}
    \includegraphics[width=\columnwidth]{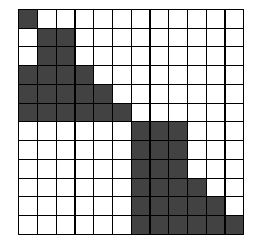}
    \caption{Multimodal packing. Several sequences with different attention patterns can be packed for more efficient processing.}
    \label{fig:attention_mask3}
  \end{subfigure}
  \caption{Various attention masks used in MLLM training.}
  \label{fig:attention_masks}
\end{figure}

We use three different types of attention patterns in our evaluation.
Figure~\ref{fig:attention_masks} illustrates these patterns.
Modality tokens can either be prepended or embedded, and several multimodal sequences can be packed into a single long sequence.
While the number of tokens for each modality is fixed as specified in Section~\ref{sec:experimental-setup}, the placement of the modality data is randomly assigned.

\section{Workload-Balanced Context Parallelism}
\label{sec:apdx-context-parallelism}

\begin{figure}[t]
  \centering
  \scriptsize
  \begin{subfigure}[t]{\columnwidth}
    \caption{Workload-balanced context parallelism with 32k sequence}
    \begin{tabular}{cc|rrrr} 
      \toprule
      \multicolumn{2}{c|}{Time (ms)} & \multicolumn{1}{c}{\begin{tabular}[c]{@{}c@{}}Causal\\CP\end{tabular}} & \multicolumn{1}{c}{\begin{tabular}[c]{@{}c@{}}Inter-GPU\\Balance\\Only\end{tabular}} & \multicolumn{1}{c}{\begin{tabular}[c]{@{}c@{}}Intra-GPU\\Balance\\Only\end{tabular}} & \multicolumn{1}{c}{\name}  \\ 
      \midrule
      \multirow{2}{*}{LLM-S} & Attn  & 66.00                       & 74.44                                                                                 & 57.65                                                                                 & 57.32                           \\
                              & Model & 1858.73                    & 1976.68                                                                              & 1731.12                                                                              & 1705.38                        \\ 
      \midrule
      \multirow{2}{*}{LLM-M} & Attn  & 113.37                      & 127.46                                                                                & 111.17                                                                                & 105.16                          \\
                              & Model & 8345.84                    & 8789.94                                                                              & 8213.21                                                                              & 8029.79                        \\ 
      \midrule
      \multirow{2}{*}{LLM-L} & Attn  & 148.75                      & 159.39                                                                                & 141.85                                                                                & 143.28                          \\
                              & Model & 29372.62                   & 30234.54                                                                             & 28767.26                                                                             & 28518.80                       \\
      \bottomrule
    \end{tabular}
  \end{subfigure}
  \begin{subfigure}[t]{\columnwidth}
    \caption{Workload-balanced context parallelism with 16k sequence}
    \begin{tabular}{cc|rrrr} 
      \toprule
      \multicolumn{2}{c|}{Time (ms)} & \multicolumn{1}{c}{\begin{tabular}[c]{@{}c@{}}Causal\\CP\end{tabular}} & \multicolumn{1}{c}{\begin{tabular}[c]{@{}c@{}}Inter-GPU\\Balance\\Only\end{tabular}} & \multicolumn{1}{c}{\begin{tabular}[c]{@{}c@{}}Intra-GPU\\Balance\\Only\end{tabular}} & \multicolumn{1}{c}{\name}  \\ 
      \midrule
      \multirow{2}{*}{LLM-S} & Attn  & 23.54                       & 24.26                                                                                 & 17.61                                                                                 & 18.23                           \\
                             & Model & 800.29                     & 802.07                                                                               & 691.22                                                                               & 705.14                         \\ 
      \midrule
      \multirow{2}{*}{LLM-M} & Attn  & 40.01                       & 40.72                                                                                 & 35.35                                                                                 & 34.47                           \\
                             & Model & 3682.71                    & 3700.22                                                                              & 3467.52                                                                              & 3491.53                        \\ 
      \midrule
      \multirow{2}{*}{LLM-L} & Attn  & 48.02                       & 50.77                                                                                 & 44.74                                                                                 & 41.41                           \\
                             & Model & 13089.76                   & 13184.58                                                                             & 12677.96                                                                             & 12628.59                       \\
      \bottomrule
    \end{tabular}
  \end{subfigure}
  \caption{Workload-balanced context parallelism with different sequence lengths.}
  \label{fig:apdx-workload-balanced-context-parallelism}
\end{figure}

Figure~\ref{fig:apdx-workload-balanced-context-parallelism} shows context parallelism results with smaller sequence lengths than 64k.
Similar patterns as in Table~\ref{tab:context_parallel} are observed.
Intra-GPU balance balances workloads of long sequences across SMs within each GPU.
While inter-GPU balance, if applied alone, is worse than context parallelism optimized for causal attention, it provides further optimized performance when combined with intra-GPU balance.

\begin{table}[t]
  \centering
  \small
  \caption{Model execution time with inter-GPU balancing + using multiple CUDA streams.}
  \label{tab:apdx-context_parallelism_model_execution_time}
  \begin{tabular}{c|rrr} 
    \toprule
    Time (ms) & \multicolumn{1}{c}{\begin{tabular}[c]{@{}c@{}}Causal\\CP\end{tabular}} & \multicolumn{1}{c}{\begin{tabular}[c]{@{}c@{}}Inter-GPU\\Balance +\\Multistreams\end{tabular}} & \multicolumn{1}{c}{\name}  \\ 
    \midrule
    LLM-S     & 5541.25                    & 5294.12                                                                                        & 4856.60                        \\ 
    \midrule
    LLM-M     & 24534.50                   & 23794.29                                                                                       & 22815.79                       \\ 
    \midrule
    LLM-L     & 77378.44                   & 76457.72                                                                                       & 74864.71                       \\
    \bottomrule
    \end{tabular}
\end{table}

\begin{figure}[t]
  \centering
  \small
  \begin{subfigure}[t]{\columnwidth}
      \includegraphics[width=\columnwidth]{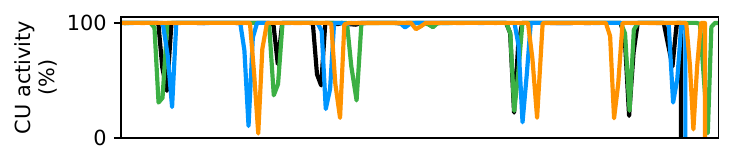}
      \caption{Small LLM}
      \label{fig:apdx-sm-analysis-multistream-s}
  \end{subfigure}

  \begin{subfigure}[t]{\columnwidth}
      \includegraphics[width=\columnwidth]{evaluations/context_parallelism/context_parallelism_sm_activity_llm_m_cornstarch_imbalanced_multistream.pdf}
      \caption{Medium LLM}
      \label{fig:apdx-sm-analysis-multistream-m}
  \end{subfigure}

  \begin{subfigure}[t]{\columnwidth}
      \includegraphics[width=\columnwidth]{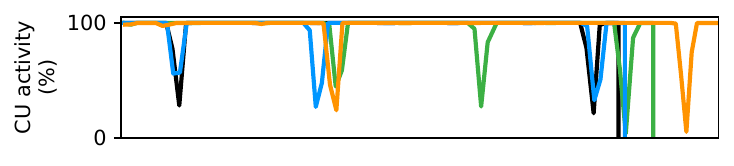}
      \caption{Large LLM}
      \label{fig:apdx-sm-analysis-multistream-l}
  \end{subfigure}
  \caption{CU activity analysis with inter-GPU balancing + multiple CUDA streams.}
  \label{fig:apdx-sm-analysis}
\end{figure}

\section{Context Parallelism Using Multiple Streams}
\label{sec:apdx-multistream}
Using multiple streams in CUDA can improve performance by allowing concurrent execution of multiple attention head computations.
It does improve performance by overlapping attention head computations in the middle of each attention layer as presented in Table~\ref{tab:apdx-context_parallelism_model_execution_time}.
However, in Figure~\ref{fig:apdx-sm-analysis}, spikes are still observed at the end of every attention iteration, as the last attention head computation cannot be overlapped with the next head.

\section{Comparison with DCP}
\label{sec:apdx-dcp-comparison}

\begin{table}[H]
  \centering
  \caption{Causal.}
  \label{tab:apdx_dcp_causal}
  \begin{tabular}{c|rrrr} 
  \toprule
  Seq length & \multicolumn{1}{c}{32k} & \multicolumn{1}{c}{64k} & \multicolumn{1}{c}{128k} & \multicolumn{1}{c}{256k}  \\ 
  \midrule
  Cornstarch & 211.51                  & 158.15                  & 380.88                   & 1189.57                   \\
  DCP        & 54.63                   & 149.18                  & 503.56                   & 1861.39                   \\
  \bottomrule
  \end{tabular}
\end{table}

\begin{table}[H]
  \centering
  \caption{Causal blockwise.}
  \label{tab:apdx_dcp_causalblockwise}
  \begin{tabular}{c|rrrr} 
  \toprule
  Seq length & \multicolumn{1}{c}{32k} & \multicolumn{1}{c}{64k} & \multicolumn{1}{c}{128k} & \multicolumn{1}{c}{256k}  \\ 
  \midrule
  Cornstarch & 228.36                  & 124.72                  & 333.91                   & 1002.70                   \\
  DCP        & 41.13                   & 104.87                  & 316.93                   & 1083.57                   \\
  \bottomrule
  \end{tabular}
\end{table}

\begin{table}[H]
  \centering
  \caption{Shared questions.}
  \label{tab:apdx_dcp_sharedquestions}
  \begin{tabular}{c|rrrr} 
  \toprule
  Seq length & \multicolumn{1}{c}{32k} & \multicolumn{1}{c}{64k} & \multicolumn{1}{c}{128k} & \multicolumn{1}{c}{256k}  \\ 
  \midrule
  Cornstarch & 199.99                  & 179.59                  & 351.41                   & 1190.41                   \\
  DCP        & 39.79                   & 120.74                  & 374.99                   & 1358.43                   \\
  \bottomrule
  \end{tabular}
\end{table}

We compare \name with DCP in terms of single attention layer performance with three different attention patterns demonstrated in the DCP paper~\cite{dcp-sosp25}: causal, causal blockwise, and shared question, as in Table~\ref{tab:apdx_dcp_causal}, Table~\ref{tab:apdx_dcp_causalblockwise}, and Table~\ref{tab:apdx_dcp_sharedquestions}.
All time is in milliseconds.

% You can have as much text here as you want. The main body must be at most $8$
% pages long. For the final version, one more page can be added. If you want, you
% can use an appendix like this one.

% The $\mathtt{\backslash onecolumn}$ command above can be kept in place if you
% prefer a one-column appendix, or can be removed if you prefer a two-column
% appendix.  Apart from this possible change, the style (font size, spacing,
% margins, page numbering, etc.) should be kept the same as the main body.
%%%%%%%%%%%%%%%%%%%%%%%%%%%%%%%%%%%%%%%%%%%%%%%%%%%%%%%%%%%%%%%%%%%%%%%%%%%%%%%
%%%%%%%%%%%%%%%%%%%%%%%%%%%%%%%%%%%%%%%%%%%%%%%%%%%%%%%%%%%%%%%%%%%%%%%%%%%%%%%

\end{document}